\pdfoutput=1
\documentclass[12pt,amsmath,amssymb,longbibliography,nofootinbib,showkeys,eprint]{revtex4-2}
\usepackage[activate={true,nocompatibility},final,tracking=true,kerning=true,spacing=true]{microtype}
\usepackage{filecontents} 
\usepackage{epigraph}
\usepackage{graphicx}
\usepackage{epstopdf}
\usepackage{braket}
\usepackage{bm}
\usepackage{numprint}
\usepackage{float}
\usepackage[T1,T2A]{fontenc}
\usepackage[utf8]{inputenc}
\usepackage{xr}
\usepackage{hhline}
\usepackage{amsthm}
\usepackage{dsfont}
\usepackage{calc}
\usepackage{seqsplit}
\usepackage{hyperref}

\allowdisplaybreaks
\begin{document}
\preprint{V.M.}
\title{Market Directional Information Derived From
  (Time, Execution Price, Shares Traded) Sequence of Transactions. \\
  On The Impact From The Future.
}
\author{Vladislav Gennadievich \surname{Malyshkin}}
\email{malyshki@ton.ioffe.ru}
\affiliation{Ioffe Institute, Politekhnicheskaya 26, St Petersburg, 194021, Russia}

\author{Mikhail Gennadievich \surname{Belov}}
\email{mikhail.belov@tafs.pro}
\affiliation{Lomonosov Moscow State University,  Faculty of Mechanics and Mathematics,
   GSP-1,  Moscow, Vorob'evy Gory, 119991, Russia}

\date{September, 20, 2022}

\begin{abstract}
\begin{verbatim}
$Id: ImpactFromTheFuture.tex,v 1.269 2022/10/09 10:41:55 mal Exp $
\end{verbatim}
An attempt to obtain market directional information
from non--stationary solution of the dynamic equation:
``future price tends to the value
maximizing the number of shares traded per unit time''
is presented.
A remarkable feature of the approach is an automatic time scale selection.
It is determined from the state of maximal execution flow
calculated on past transactions.
Both lagging and advancing prices are calculated.
\end{abstract}

\maketitle
\newpage

\epigraph{Времена Пугачёвского бунта.
Самозванец выступает перед народом,
говорит о грядущем счастье,
которое придёт в форме мужицкого царства.
Пленный офицер спрашивает: ``Откуда деньги будут на всю эту благодать''?
Пугачёв ответил: ``Ты что, дурак? Из казны жить будем!''}{Народная легенда, 1774.}

\section{\label{intro}Introduction}

Introduced in \cite{ArxivMalyshkinMuse}
the ultimate market dynamics problem: an evidence of existence (or a
proof of non--existence) of an automated trading machine,
consistently making positive P\&L trading on a free market as an autonomous agent
can be formulated in its weak and strong forms\cite{MalMuseScalp}:
whether such an automated trading machine
can exist with legally available data (weak form)
and
whether it
can exist with transaction sequence triples
(time, execution price, shares traded) as the only information available (strong form); in the later case
execution flow $I=dV/dt$ is the only available characteristic
determining market dynamics.

Let us formulate the problem in the third, ``superstrong'', form:
Whether the future value of price
can be predicted
from
  (time, execution price, shares traded) sequence of past transactions?
Previously\cite{2015arXiv151005510G,2016arXiv160305313G}
we thought this is not possible,
only P\&L that includes not only price dynamics but also trader actions
can be possibly predicted. Recent results changed our opinion.

There are  two types of predicted price: ``lagging'' (retarded)  and ``advancing'' (future)
Lagging price $P^{Ret}$ corresponds to past observations;
future direction is determined by the difference of last price
$P^{last}$ and $P^{Ret}$. An example of $P^{Ret}$ is moving average.
A common problem with lagging price is that it typically assumes
an existence of a time scale the $P^{Ret}$ is calculated with,
what gives incorrect direction 
for market movements with time scales lower than the one of $P^{Ret}$;
however making the time scale too low creates a large amount of false signals.
Advancing price $P^{Adv}$ is predicting actual value of future price;
the direction is determined by the difference of
$P^{Adv}$ and $P^{last}$. The $P^{Adv}$ is typically calculated from
limit order book information, brokerage clients order flow timings, etc.

In this work both lagging and advancing prices
are calculated
from
(time, execution price, shares traded) sequence of past transactions.
The key element is to determine the state $\Ket{\psi^{[IH]}}$
of maximal execution flow $I=dV/dt$
(eigenvalue problem (\ref{GEVdef})),
as experiments show it's importance
for market dynamics. Found $\Ket{\psi^{[IH]}}$ state automatically selects the
time scale what makes the approach robust.

Found lagging price (\ref{PEQVFormula})
is the price in $\Ket{\psi^{[IH]}}$ state $P^{[IH]}$ plus trending term
that suppresses false signals.
The advancing price is obtained by
considering density matrix state $\|\rho_{JIH}\|$
corresponding to the state ``since $\Ket{\psi^{[IH]}}$ till now''
and experimentally observed fact that
operators $\left\|p\frac{dI}{dt}\right\|$ and
$\left\|I\frac{dp}{dt}\right\|$ have to be equal in $\|\rho_{JIH}\|$ state.
This corresponds to the result of our previous works
\cite{2015arXiv151005510G,2016arXiv160204423G}:
execution flow $I=dV/dt$ (the number of shares traded per unit time),
not trading volume $V$ (the number of shares traded),
is the driving force of the market: asset price is
much more sensitive to execution flow $I$ (dynamic impact),
rather than to traded volume $V$ (regular impact).

This paper is concerned only with obtaining
directional information from a sequence of past transaction
in a ``single asset universe'' just for simplicity,
see Section \ref{conclusion} below for multi asset universe generalization.
Whereas the dynamics theory of Section \ref{impactFromTheFuture}
definitely requires additional research,
the lagging indicator (\ref{PEQVFormula}) of Section \ref{DirectionInfo},
see Fig. \ref{PVforVsurrogate},
can be practically applied to trading even in a single asset universe.
In this work we do not implement any trading ideas of
\cite{2015arXiv151005510G,2016arXiv160305313G},
where a concept of liquidity deficit trading:
\textsl{open a position at low $I$, then
close already opened position at high $I$},
as this is the only strategy that avoids eventual catastrophic  P\&L losses.
This paper is concerned only with obtaining a directional information
that is required to  determine
what side the position has to be open
on a liquidity deficit event.

\section{\label{StateMaxI}The State Of Maximal Execution Flow}
Introduce a wavefunction $\psi(x)$ as a linear combination
of basis function $Q_k(x)$:
\begin{eqnarray}
  \psi(x)&=&\sum\limits_{k=0}^{n-1} \alpha_k Q_k(x) \label{psiintr}
\end{eqnarray}
Then an observable market--related value
$f$, corresponding to probability density  $\psi^2(x)$,
is calculated by averaging timeserie sample
with the weight $d\mu=\psi^2(x(t))\omega(t)dt$;
the expression corresponds to an estimation of Radon--Nikodym
derivative\cite{malyshkin2019radonnikodym}:
\begin{eqnarray}
  f_{\psi}&=&\frac{\Braket{\psi|f|\psi}}{\Braket{\psi|\psi}}
  \label{faver} \\
  f_{\psi}&=&\frac{\sum\limits_{j,k=0}^{n-1}\alpha_j\Braket{Q_j|f|Q_k}\alpha_k}{\sum\limits_{j,k=0}^{n-1}\alpha_j\Braket{Q_j|Q_k}\alpha_k}
  \label{faverexpand}
\end{eqnarray}
For averages we use
\href{https://en.wikipedia.org/wiki/Bra%E2%80%93ket_notation}{bra--ket notation}
  by
  \href{https://en.wikipedia.org/wiki/Paul_Dirac}{Paul Dirac}:
  $\Bra{\psi}$ and $\Ket{\psi}$.
The (\ref{faver}) is plain ratio of two moving averages,
but the weight is not regular decaying exponent $\omega(t)$
from (\ref{Wbasis}),
but exponent multiplied by wavefunction squared as $d\mu=\psi^2(x(t))\omega(t)dt$,
the $\psi^2(x)$ defines how to average a timeserie sample.
Any $\psi(x)$ function is defined by $n$ coefficients $\alpha_k$,
the value of an observable variable $f$ in $\psi(x)$ state
is a ratio of
two quadratic forms on $\alpha_k$ (\ref{faverexpand});
as an example of a wavefunction see localized state (\ref{psi0}),
it can be used for Radon--Nikodym interpolation:
$f(y)\approx\Braket{\psi_y|f|\psi_y}\Big/\Braket{\psi_y|\psi_y}$;
familiar least squares interpolation
is  also available:
$f(y)\approx\Braket{\psi_y|f}\psi_y(y)=\sum_{j,k=0}^{n-1}\Braket{Q_jf}G^{-1}_{jk}Q_k(y)$.

One can also consider a more general form of average,
$d\mu=P(x(t))\omega(t)dt$,
where $P(x)$ is an arbitrary polynomial, not just the square
of a wavefunction. These states correspond to a density matrix average:
\begin{eqnarray}
  f_{\rho_P}&=&\frac{\mathrm{Spur}\left\|f\middle|\rho_P\right\|}{\mathrm{Spur}\left\|\rho_P\right\|}
  \label{faverDM}
\end{eqnarray}
This average, the same as (\ref{faver}), is a ratio of two moving averages.
For an algorithm to convert a polynomial $P(x)$ to the
density matrix $\|\rho_P\|$
see Theorem 3 of \cite{ArxivMalyshkinLebesgue}.
A useful application of the density matrix states is to study
an average ``since $\Ket{\psi}$''; for example
if $\Ket{\psi}$ corresponds to a past $dV/dt$ spike, then the polynomial
``since $\Ket{\psi}$ till now''
is
$P(x)=J(\psi^2(x))$ with $J(\cdot)$ defined in (\ref{JPDforDifferentBases});
price change between ``now'' and the time of spike
is $P^{last}-\Braket{\psi|p|\psi}=\mathrm{Spur}\left\|\frac{dp}{dt}\middle|\rho_{J(\psi^2)}\right\|$,
similarly, total traded volume on this interval
is $\mathrm{Spur}\left\|\frac{dV}{dt}\middle|\rho_{J(\psi^2)}\right\|$.

The main idea of \cite{2015arXiv151005510G}
is to consider a wavefunction (\ref{psiintr})
then to construct (\ref{faverexpand}) quadratic forms ratio.
A
\href{https://en.wikipedia.org/wiki/Eigendecomposition_of_a_matrix\#Generalized_eigenvalue_problem}{generalized eigenvalue problem}
can be considered with the two matrices from (\ref{faverexpand}).
The most general case corresponds to \textsl{two} operators
$A$ and $B$. Consider an eigenvalue problem
with the matrices $\Braket{Q_j|A|Q_k}$ and $\Braket{Q_j|B|Q_k}$:
\begin{align}
  \Ket{A\middle|\psi^{[i]}}&=\lambda^{[i]}\Ket{B\middle|\psi^{[i]}} \label{GEVgeneral}\\
\sum\limits_{k=0}^{n-1} \Braket{Q_j|A|Q_k} \alpha^{[i]}_k &=
  \lambda^{[i]} \sum\limits_{k=0}^{n-1} \Braket{ Q_j|B|Q_k} \alpha^{[i]}_k
  \label{GEVgeneralMatrix} \\
  \psi^{[i]}(x)&=\sum\limits_{k=0}^{n-1} \alpha^{[i]}_k Q_k(x) \label{psiDefAB} \\
  \delta_{ij}&=\Braket{\psi^{[i]}|B|\psi^{[j]}}
  =\sum\limits_{k,m=0}^{n-1}\alpha^{[i]}_k \Braket{ Q_k|B|Q_m} \alpha^{[j]}_m
  \label{normPsi} \\
  \lambda^{[i]}\delta_{ij}&=\Braket{\psi^{[i]}|A|\psi^{[j]}}
  =\sum\limits_{k,m=0}^{n-1}\alpha^{[i]}_k \Braket{ Q_k|A|Q_m} \alpha^{[j]}_m
  \label{normPsiEV}
\end{align}
If at least one
of these two matrices is positively defined -- the problem
has a unique solution (within eigenvalues degeneracy).
In the found basis $\Ket{\psi^{[i]}}$ the two matrices
are simultaneously diagonal:
(\ref{normPsi}) and (\ref{normPsiEV}).
See (\ref{MatrFromBToQ}) to convert an operator's matrix from
$\Ket{\psi^{[i]}}$ to $Q_j$ basis and (\ref{MatrFromQToB})
to convert it from $Q_j$ to $\Ket{\psi^{[i]}}$ basis.

In our previous work
\cite{2015arXiv151005510G,2016arXiv160204423G,ArxivMalyshkinMuse,MalMuseScalp}
we considered various $A$ and $B$ operators,
with the goal to find operators and states that are related to market dynamics.
We established,
that execution flow $I=dV/dt$ (the number of shares traded per unit time),
not trading volume $V$ (the number of shares traded),
is the driving force of the market: asset price is
much more sensitive to execution flow $I$ (dynamic impact),
rather than to traded volume $V$ (regular impact).
This corresponds to the matrices
$\Braket{Q_j|I|Q_k}=\Braket{Q_j|A|Q_k}$ and $\Braket{Q_j|Q_k}=\Braket{Q_j|B|Q_k}$.
These two matrices are volume- and time- averaged products of two basis functions.
Generalized eigenvalue problem for operator $I=dV/dt$
is the equation to determine market dynamics:
\begin{align}
&\Ket{I\middle|\psi^{[i]}}=\lambda^{[i]}\Ket{\psi^{[i]}} \label{GEVdef} \\
&\sum\limits_{k=0}^{n-1} \Braket{Q_j|I|Q_k} \alpha^{[i]}_k =
  \lambda^{[i]} \sum\limits_{k=0}^{n-1} \Braket{ Q_j|Q_k} \alpha^{[i]}_k
  \label{GEV} \\
 \psi^{[i]}(x)&=\sum\limits_{k=0}^{n-1} \alpha^{[i]}_k Q_k(x)
 \label{psiC}\\
 \psi_y(x)&=\frac{\sum\limits_{i=0}^{n-1} \psi^{[i]}(y)\psi^{[i]}(x)}
     {\sqrt{\sum\limits_{i=0}^{n-1} \left[\psi^{[i]}(y)\right]^2}}
 =\sum\limits_{i=0}^{n-1} \Ket{\psi^{[i]}}\Braket{\psi_y|\psi^{[i]}}
 =\frac{\sum\limits_{j,k=0}^{n-1}Q_j(x)G^{-1}_{jk}Q_k(y)}
  {\sqrt{\sum\limits_{j,k=0}^{n-1}Q_j(y)G^{-1}_{jk}Q_k(y)}}
\label{psi0}\\
\Braket{\psi_{y}|\psi^{[i]}}^2&=\left[\frac{\psi^{[i]}(y)}{\psi_{y}(y)}\right]^2
=\frac{\left[\psi^{[i]}(y)\right]^2}{\sum\limits_{k=0}^{n-1} \left[\psi^{[k]}(y)\right]^2}
\label{proj}
\end{align}
The $y=x_0$ is the time ``now'',
$\psi_y(x)$ is a wavefunction localized at $x=y$.
Here and below we write $\psi_0(x)$ instead of $\psi_{x_0}(x)$
to simplify notations.
The $\Braket{\psi_0|\psi^{[i]}}$ is the projection
of the $\Ket{\psi^{[i]}}$ state of (\ref{GEVdef}) eigenproblem
to the state ``now'' $\Ket{\psi_0}$.

Our analysis\cite{2015arXiv151005510G,2016arXiv160204423G,ArxivMalyshkinMuse,MalMuseScalp}
shows that among the states $\Ket{\psi^{[i]}}$ of
the problem (\ref{GEVdef})
the state corresponding
to the maximal eigenvalue among all $\lambda^{[i]}$, $i=0\dots n-1$,
is the most important for market dynamics.
Consider various observable characteristics
in this state $\Ket{\psi^{[IH]}}$.

\begin{figure}[t]
  \includegraphics[width=16cm]{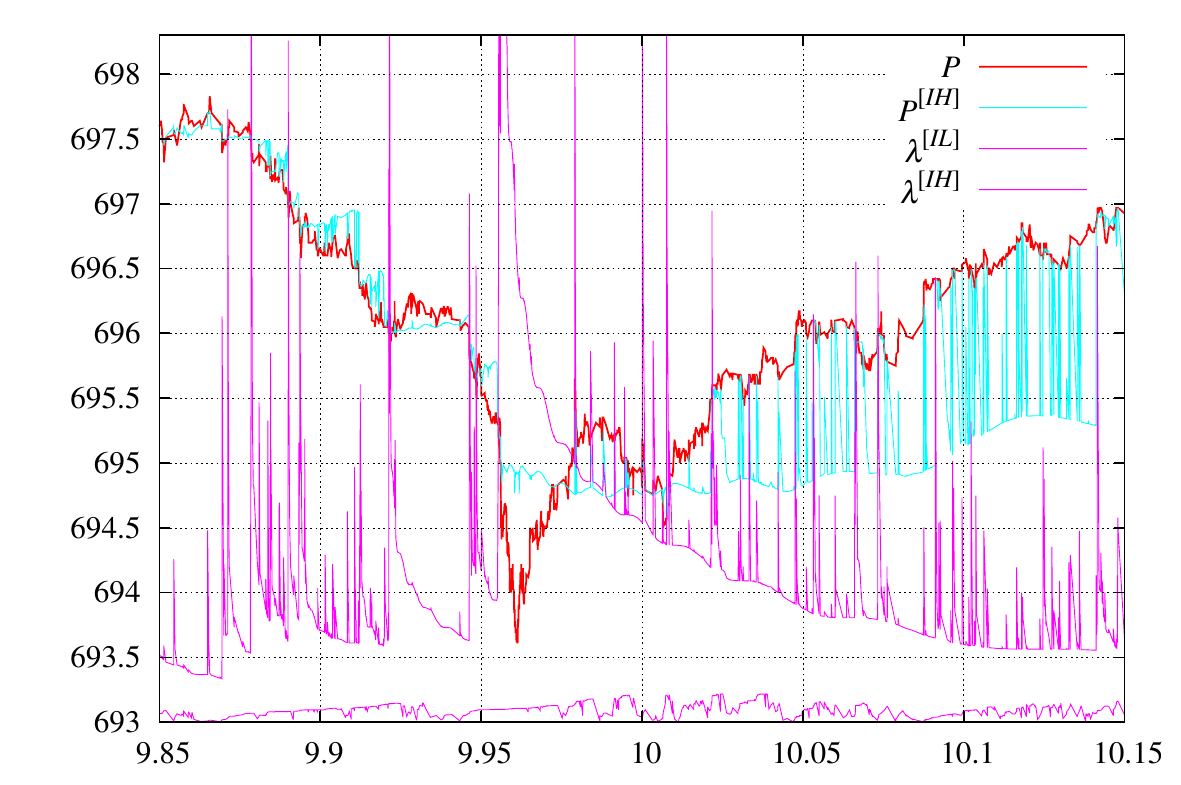}
  \caption{\label{ExampleIPsiH}
    Price $P$, price $P^{[IH]}$ (\ref{PIHGEV}),
    and maximal/minimal eigenvalues of (\ref{GEVdef})
    for
 AAPL stock on September, 20, 2012.
 The calculations in shifted Legendre basis with $n=12$ and $\tau$=128sec.
  The execution flow  eigenvalues are scaled and shifted to $693$ to fit the chart.
  }
\end{figure}

In Fig. \ref{ExampleIPsiH} a demonstration of
several observables:
the price in $\Ket{\psi^{[IH]}}$ state (\ref{PIHGEV}),
maximal eigenvalue $\lambda^{[IH]}$ of (\ref{GEVdef})   problem,
and minimal eigenvalue $\lambda^{[IL]}$ (for completeness)
are presented.
\begin{align}
  P^{[IH]}&=\frac{\Braket{\psi^{[IH]}|pI|\psi^{[IH]}}}{\Braket{\psi^{[IH]}|I|\psi^{[IH]}}} \label{PIHGEV}
\end{align}
From these observable one can clearly see
that singularities in $I$ cause singularities in price,
and that a change in $\Ket{\psi^{[IH]}}$ localization causes
an immediate ``switch'' in an observable.
This switch is caused by the presence
of $n-1$ internal degrees of freedom
$\alpha_k$ ($n$ coefficients, one less due to
normalizing $1=\Braket{\psi|\psi}$, Eq.  (\ref{normPsi})).
Such a ``switch'' is not possible in regular moving average
(\ref{pmovingaver})
since it has no any internal degree of freedom,
hence, all regular moving average dependencies are smooth.

The state  $\Ket{\psi^{[IH]}}$
that maximizes the number of shares traded per unit
time on past observations sample
is the main result of our initial work
\cite{2015arXiv151005510G}.

\section{\label{TimeScale}On Time Scale Selection of a Trading Strategy }
Financial markets have no intrinsic time scales\footnote{
Trivial time scales such as seasonal, daily open/close, year end, etc.
while actually do exist provide little trading opportunities.
}
(at least those a market participant can take an advantage of).
For US equity market --- market timeserie data manifests
an existence of time scales from microseconds to decades.
For NASDAQ ITCH \cite{itchfeed} data feed time-discretization is one nanosecond.
Whereas real markets typically have no intrinsic time scale,
any trading strategy typically does have an intrinsic time scale.
This time scale is determined by:
available data feeds, available execution, trader personal preferences, etc.
An implementation of trading strategies with
time scales under one second requires a costly IT infrastructure of data feed/execution,
and is hard to program algorithmically; moreover,
market liquidity at such a low time scale is low, a
situation when a dozen of HFT firms are chasing
a single limit order of 100 shares is very common.
For trading strategies with a large time scale the major difficulty
is that a trader, observing post-factum missed opportunities,
often starts to ``adjust'' the strategy to lower time scales.
For professional money managers (managing other people money),
with the rare exception of ``super-stars'',
the maximal possible time scale is one month: once a month
a letter to investors explaining the fund performance is required to be sent.
There is no such a ``monthly'' constraint for somebody managing his own money,
for example, an individual crypto investor may be 50\% down
in April 2022 -- but for him this problem is not as big as it were
for a fund. For traders the most popular time scales
are between ``daily trading'' and ``monthly P\&L''; these time scales provide
sufficient number of opportunity events along with
market data availability (e.g. Bloomberg).
Important that these time scales are ``compatible'' with human reaction time.

The major drawbacks of trading systems the authors observed
among institutional investors/hedge funds/individual investors
is that all of them typically have a few time scales.
Most often -- a single time scale.
It may be explicit or implicit, but it almost always exists.
The contradiction between a spectrum of time scales of the financial markets
and a single time scale of a trading system is
the most common limitation in trading systems design.

\begin{figure}[t]
  \includegraphics[width=16cm]{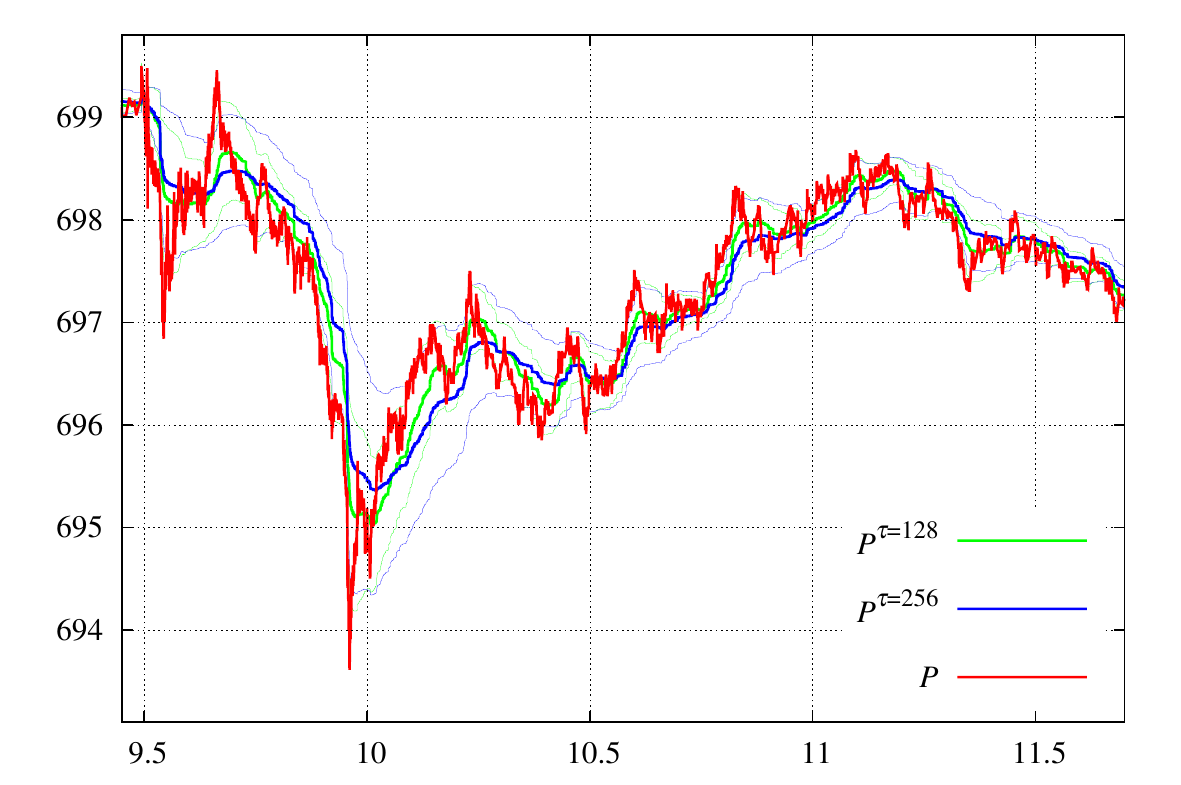}
  \caption{\label{MovingAveragePlot}
    An example of regular exponential moving average
    corresponding to $\tau=128$s and $\tau=256$s.
    Standard deviation is also calculated with the same $\tau$ and
    moving average $\pm$ standard deviation is plotted as a thin line in the same color. As $\tau$ increases -- the moving average ``shifts to the right''
    ($\tau$-proportional time delay, lagging indicator).
 The data is for AAPL stock on September, 20, 2012.
  }
\end{figure}

Consider familiar demonstration with
a moving average. Let $P^{\tau}$ be a regular exponential moving average.
The average $\Braket{\cdot}$
is calculated with the weight (\ref{Wbasis}):
\begin{align}
  P^{\tau}(t_{now})&=\frac{\Braket{pI}}{\Braket{I}}
  =\frac{\Braket{Q_0pI}}{\Braket{Q_0I}}
  =\frac{\Braket{Q_0|pI|Q_0}}{\Braket{Q_0|I|Q_0}}
  =\frac{\int_{-\infty}^{t_{now}} dV\,\omega(t)p(t)}{\int_{-\infty}^{t_{now}} dV\,\omega(t)}
  \label{pmovingaver}
\end{align}
The averaging $d\mu=\omega(t)dt$ takes place between the past and $t_{now}$
using exponentially decaying weight $\omega(t)=\exp\left(-(t_{now}-t)/\tau\right)$.
With $\tau$ increase, the contributing to integral interval becomes larger
and moving average ``shifts to the right'' ($\tau$-proportional time delay, lagging indicator).
The (\ref{pmovingaver}) has no single parameter that can ``adjust''
the time scale as $\alpha_k$ do in (\ref{faverexpand})
where $d\mu=\psi^2(x(t))\omega(t)dt$.
In (\ref{pmovingaver}) we have $\psi(x)=const$.
Trading strategies that watch crossing between price
and moving average,
or between two moving averages calculated with different values of $\tau$,
have the problem that the specific values of time scales are initially preset.
We personally observed a number of successful (and failed) traders
who were constantly watching
moving averages on Bloomberg --- we may tell that their success
is caused by intuitively
switching from one scale to another;
if you ask such a person what he is doing -- he cannot explain;
but looking at him from a side it is clear -- the person is trying
to identify relevant time- and price- scales.
Successful traders also jump frequently
by observing assets of different classes; it is a common situation
before placing a trade on GOOG
to observe: DJI, AAPL, commodity, power generating industry, chemical industry --
all withing less than a minute.
If you go from a human
(who select the time scale based on intuition, market knowledge, news,
personal communications, experience, etc.)
to an ``automated trading machine'' that has none of that -- the problem
of selecting the time scale becomes very difficult.
The problem of automatic time scale selection
is crucial in trading systems design.
Another critically important problem is to adsorb information
of different financial instruments. The theory presented below
is perfectly applicable in multi asset universe; the analysis and interpretation,
however,
become more complicated, see Section \ref{conclusion}
below for a discussion.
In this paper we will be concerned only with a single asset
universe to demonstrate the main ideas, and a detailed generalization
of the theory to multi asset universe will be published elsewhere.

Whereas we still have no approach to price scale selection
(the \cite{2016arXiv160204423G} uses price basis $p^k$ as $Q_k$,
but in the stationary case this is actually a time scale equivalent),
we do have a practical method for an authomatic
selection of the time scale.

Considered in Section \ref{StateMaxI} above
the state  $\Ket{\psi^{[IH]}}$
that maximizes the number of shares traded per unit
time on past observations sample
determines the time scale.
Let us consider in this state not the price and execution flow
as we studied before, but simply time distance to ``now'' in $\Ket{\psi^{[IH]}}$ state:
\begin{align}
  T^{[IH]}&=\frac{\Braket{\psi^{[IH]}|(t_{now}-t)I|\psi^{[IH]}}}{\Braket{\psi^{[IH]}|I|\psi^{[IH]}}} \label{TIHGEV} \\
  T^{\tau}&=\frac{\Braket{(t_{now}-t)I}}{\Braket{I}} \label{TregularMovingAvergage}
\end{align}
here $T^{\tau}$ is regular moving average.
As all the values of time (future and past)
are known, the (\ref{TIHGEV}) carry information
about $\Ket{\psi^{[IH]}}$ localization.
When the value is small -- a large $dV/dt$ spike event happened
very recently. When it is large -- a large spike happened a substantial time ago,
the value is an information when a large spike in $dV/dt$ took place.

\begin{figure}[t]

  \includegraphics[width=12.5cm]{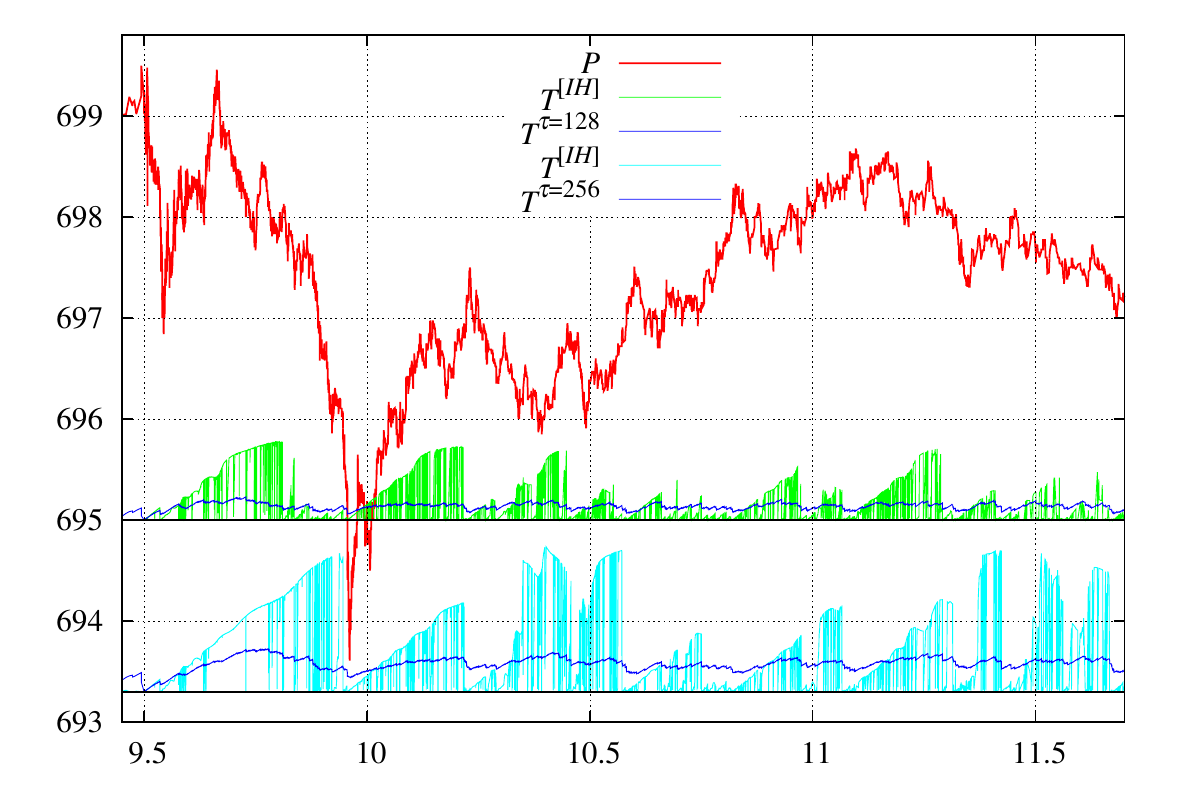}
  \includegraphics[width=12.5cm]{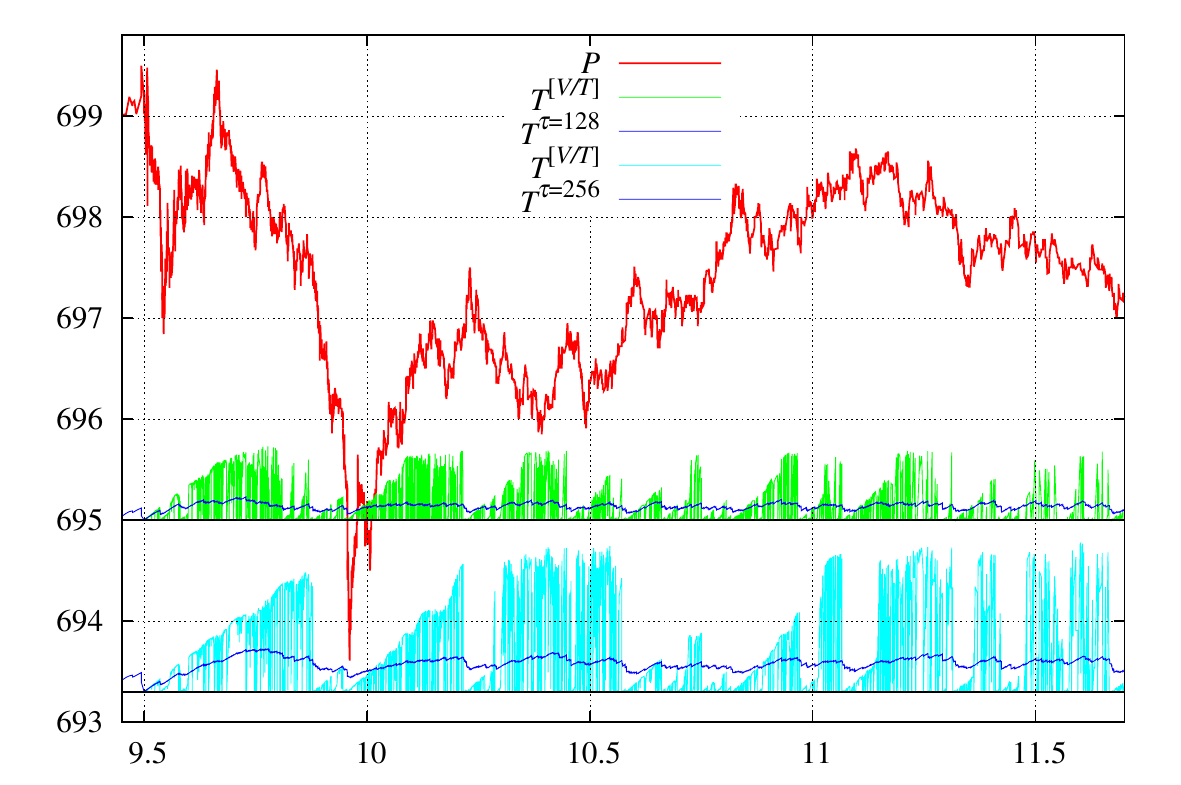}
  \caption{\label{TforImax}
    $T^{[IH]}$
    and regular moving average (\ref{TregularMovingAvergage})
    (dark blue) $T^{\tau}$ for $\tau=128$s and $\tau=256$s;
    the values are multiplied by $10^{-3}$ and shifted up to fit the chart.
 The AAPL stock on September, 20, 2012.
 The calculations in shifted Legendre basis with $n=12$.
    Top: for operator $I=dV/dt$.
    Bottom: for operator $V/T$.
  }
\end{figure}

In Fig. \ref{TforImax} (top) the value of $T^{[IH]}$ (scaled by the factor $10^{-3}$
and shifted up to fit the chart) is presented
for $\tau=128$s and $\tau=256$s.
One can clearly see that there is no smooth transition
between the states, the ``switch'' happens instantly,
there is no $\tau$-proportional time delay,
what is typical for regular moving averages $T^{\tau}$.
A linear dependence of $T^{[IH]}$ on time is also observed, this is an indication
of stability of $\Ket{\psi^{[IH]}}$ state identification.
The value of $T^{[IH]}$ is the time scale; typically it is
easier to work with
the density matrix $\rho_{J(\psi^2)}$
obtained from
$\psi(x)=\psi^{[IH]}(x)$
rather than with the time scale itself;
a typical operation with time scale -- calculate an average of some
observable in the interval of time scale length till ``now'':
the density matrix does exactly this.

We have tried a number of other operator pairs
in generalized eigenvalue problem (\ref{GEVgeneralMatrix})
\begin{center}
\begin{tabular}{||>{\raggedright}p{5cm}|p{3cm}|p{6.5cm}||}
\hhline{|t:=:=:=:t|}
$A$ & $B$ & eigenvalue meaning \\[0.5ex]
\hhline{||-|-|-||}
$\Braket{Q_j|I|Q_k}$ & $\Braket{Q_j|Q_k}$ & $I=dV/dt$ (execution flow)\\
$\Braket{Q_j|V|Q_k}$ & $\Braket{Q_j|T|Q_k}$ & $V/T$ (aggregated execution flow)\\
$\Braket{Q_j|TI-V|Q_k}$ & $\Braket{Q_j|T|Q_k}$ & $I-V/T$ \\
$\Braket{Q_j|V|Q_k}$ & $\Braket{Q_j|Q_k}$ & $V$ (traded volume)\\
$\Braket{Q_j|p|Q_k}$ & $\Braket{Q_j|Q_k}$ & $p$ (price)\\
$\Braket{Q_j|pI|Q_k}$ & $\Braket{Q_j|I|Q_k}$ & $p$ (price)\\
$\Braket{Q_j|\frac{dp}{dt}|Q_k}$ & $\Braket{Q_j|Q_k}$ & $dp/dt$\\
$\Braket{Q_j|\frac{dp}{dt}|Q_k}$ & $\Braket{Q_j|I|Q_k}$ & $dp/dV$ (market impact)\\
$\Braket{Q_j|P^{last}-p|Q_k}$ & $\Braket{Q_j|V|Q_k}$ & $\frac{P^{last}-p}{V}$ (aggregated market impact)\\
\dotfill & \dotfill & \dotfill \\
\hhline{|b:=:=:=:b|}
\end{tabular}
\end{center}
among many others\cite{ArxivMalyshkinMuse,MalMuseScalp}.
An eigenproblem with
 an additional constraint was also considered,
 see ``Appendix F'' of \cite{MalMuseScalp}
 and, more generally, ``Appendix G'' of \cite{malyshkin2019radonnikodym}.
All price-related operators cause noisy behavior,
no ``switching'' whatsoever.
Only the operator $V/T$ does
have similar to $I=dV/dt$ switching (but less pronounced); it is also
more sensitive to $\tau$ selection.
Time to max spike in $V/T$ is presented in Fig. \ref{TforImax} (bottom).
See \cite{MalMuseScalp}
about the properties of $\Ket{V|\psi}=\lambda\Ket{T|\psi}$ states:
``Appendix C: The state of maximal aggregated execution flow $V/T$''.

This makes us to conclude
that the state to determine the time scale is
the state  $\Ket{\psi^{[IH]}}$
that maximizes the number of shares traded per unit
time on past observations sample.
This state allows us to average an observable $f$
with the weights:
\begin{center}
\begin{tabular}{||>{\raggedright}p{3.8cm}|p{7.9cm}|p{4.3cm}||}
\hhline{|t:=:=:=:t|}
meaning & measure &  $\int f d\mu$ \\[0.5ex]
\hhline{||-|-|-||}
``at spike'' &
$d\mu={\psi^{[IH]}}^2(x(t))\omega(t)dt$
& $\Braket{\psi^{[IH]}|f|\psi^{[IH]}}$ \\
variation ``at spike'' &
$d\mu=2\,\mathrm{ED}\left(\psi^{[IH]}\right)\psi^{[IH]}(x(t))\omega(t)dt$
& $2\Braket{\mathrm{ED}\left(\psi^{[IH]}\right)|f|\psi^{[IH]}}$ \\
``since spike till now'' &
$d\mu=P(x(t))\omega(t)dt$; \hfill
$P(x)=J\left({\psi^{[IH]}}^2\right)$ &
$\mathrm{Spur}\|f|\rho_P\|$ \\
``since since spike'' &
$d\mu=P(x(t))\omega(t)dt$;\hfill
$P(x)=J\left(J\left({\psi^{[IH]}}^2\right)\right)$ &
$\mathrm{Spur}\|f|\rho_P\|$ \\
\hhline{|b:=:=:=:b|}
\end{tabular}
\end{center}
Found solution automatically adjusts averaging
weight
what makes the value of parameter $\tau$ in (\ref{Wbasis}) much less important.
The ``switch'' happens instantly,
without a $\tau$-proportional time delay
as it were for a regular moving average.

\section{\label{impactFromTheFuture} On The Impact From The Future}
The concept of the \textsl{Impact From The Future}
was introduced in \cite{ArxivMalyshkinMuse}.
It predicts the value of future execution flow.
Given currently observed (at $t=t_{now}$) value
of execution flow $I_0=\Braket{\psi_0|I|\psi_0}$
we know with certainty that future value of execution
flow $I_0^F$ will be greater than $I_0$ because more trading will
definitely occur in the future.
But how to estimate the value of $I_0^F$?
The
maximal eigenvalue $\lambda^{[IH]}$ of (\ref{GEVdef})
is used as the estimation of future execution flow $I_0^F$:
\begin{align}
  I_0^{F}&=\lambda^{[IH]} \label{iofuture} \\
  dI^{F} &= I_0^{F} -I_0  \label{dI}  \\
  dI^{F} &\ge 0 \label{dIge0}
\end{align}
Whereas the $I_0$ is an ``impact from the past'' (already observed current execution flow),
the $dI^{F}$ is an ``impact from the future'' (not yet observed contribution
to current execution flow);
it's value is non--negative by construction.
Similar ideology  (use past maximal value as an estimator of future value)
is often applied by market practitioners
to asset prices or their standard deviations.
This is incorrect.
Experimental observations
  show: this ideology
  is applicable {\em only} to execution flow $I=dV/dt$,
  not to the trading volume, asset price standard deviation
or any other observable.

A criterion of no information about the future can be formulated.
If current $I_0$
  is close to $\lambda^{[IH]}$, this means that we have
  a ``very dramatic market'' right now and there is no information
  about the future of this market:
\begin{align}
    dI^{F}&=0
    \label{dIeq0}
\end{align}
An alternative form of (\ref{dIeq0}) is more convenient
in practice because the value is $[0:1]$ bounded:
\begin{align}
    \Braket{\psi_0|\psi^{[IH]}}^2&=1
    \label{dIeq0Psi}
\end{align}
This means that $\Ket{\psi^{[IH]}}$ and $\Ket{\psi_0}$ are the same
(\ref{proj}).
In practice a good value of the threshold is between $[0.2:0.8]$
instead of the maximal value of $1$.
In Fig. \ref{ExampleIPsiH} one can clearly see
the spikes in $\lambda^{[IH]}$ when (\ref{dIeq0Psi}) approaches $1$,
for $I_0(t)$ see Fig. 1 of \cite{MalMuseScalp}, it is not
presented in Fig. \ref{ExampleIPsiH} to save the place.

We have the state $\Ket{\psi^{[IH]}}$ and the criterion (\ref{dIeq0Psi}) of
no information about the future.
How to obtain directional information?
Previously \cite{2015arXiv151005510G} we considered the price $P^{[IH]}$
(\ref{PIHGEV}) in the found state $\Ket{\psi^{[IH]}}$
as an indicator related to market direction. The difference between
$P^{last}$ and $P^{[IH]}$ was used as a directional indicator.
A typical
result is presented in Fig. \ref{ExampleIPsiH}. The price $P^{[IH]}$
is actually a moving average with positive weight having $n-1$ internal
degrees of freedom. It determines the direction (and can possibly work for ``reverse to the mean'' type of strategy), but this is not the future price.

Consider a concept from classical machanics. Let us introduce a Lagrangian--like
function:
\begin{align}
  {\cal S}&=\int\limits_{t_1}^{t_2}L(p,V,t)dt
  \label{lagr}
\end{align}
and try to variate it. Let us first take $L$ to be an exact differential
(e.g. total energy in classical mechanics $L=T+U$). Then (\ref{lagr})
carry no information about the dynamics, but we obtain two
distinct
terms $\int_{t_1}^{t_2}Tdt$ and $\int_{t_1}^{t_2}Udt$
that we can consider the difference of and obtain actual equation of motion.
We implemented this strategy below by considering various hypothesises
for action $\cal S$ and testing them experimentally.

\subsection{\label{VolumeBasedDynamics}Volume Driven Dynamics}
Assume that price changes are caused by trading volume.
Introduce
\begin{align}
  L(p,V)&=\frac{d}{dt}(p-P^{last})(V-V^{last})\nonumber\\
  &=(V-V^{last})\frac{dp}{dt}+(p-P^{last})\frac{dV}{dt}
  \label{lagrangianV}
\end{align}
Then ``exact differential action''
\begin{align}
  {\cal S}^{ed}&=
    \Braket{\psi^{[IH]}\left|(V-V^{last})\frac{dp}{dt}\right|\psi^{[IH]}}    
+
    \Braket{\psi^{[IH]}\left|(p-P^{last})I\right|\psi^{[IH]}}
    \label{SVnodynamics}
\end{align}
To obtain ``actual'' action we have to change the sign in between:
\begin{align}
  {\cal S}&=
    \Braket{\psi^{[IH]}\left|(V-V^{last})\frac{dp}{dt}\right|\psi^{[IH]}}    
-
    \Braket{\psi^{[IH]}\left|(p-P^{last})I\right|\psi^{[IH]}}
    \label{SV}
\end{align}
These two terms can be considered as ``kinetic'' and ``potential'' energy.
It is difficult to variate (\ref{SV})
so let us just find the price $P^{last}$
that makes these two terms equal\footnote{Similar concept
in mechanics corresponds to finding the state in which kinetic energy equals
to potential energy. This gives an exact result for oscillators and approximate
result for many other systems, see
\href{https://en.wikipedia.org/wiki/Virial_theorem}{Virial theorem}.}:
\begin{align}
  \Delta_{VD}&=\Braket{\psi^{[IH]}\left|(V-V^{last})\frac{dp}{dt}\right|\psi^{[IH]}}
  \label{DeltaVolume} \\
  P^{EQ}&=P^{[IH]}-  \frac{1}{\lambda^{[IH]}}\Delta_{VD}
  \label{PEQV}
\end{align}
The dynamics with (\ref{SV}) action is a volume-driven dynamics.
Let $I=\frac{dV}{dt}=const$ then both price and volume
are linear functions on time,
$P=\alpha t + C$ and $V-V^{last}= It$, and two terms in (\ref{SV}) are equal
exactly in any state:
$(p-P^{last})I=(V-V^{last})\frac{dp}{dt}$.
Taking into account that $\Ket{\psi^{[IH]}}$ is typically localized
(see Fig. 6 of Ref. \cite{MalMuseScalp})
obtain
$\Braket{\psi^{[IH]}\left|(V-V^{last})\frac{dp}{dt}\right|\psi^{[IH]}}
\approx
\Braket{\psi^{[IH]}|V-V^{last}|\psi^{[IH]}}\Braket{\psi^{[IH]}|\frac{dp}{dt}|\psi^{[IH]}}$.
Since $V-V^{last}$ is negative, the (\ref{PEQV})
means trend following, the trend is determined in $\Ket{\psi^{[IH]}}$ state.
In this state we have price equals to $P^{[IH]}$
and $dp/dt=\Braket{\psi^{[IH]}|\frac{dp}{dt}|\psi^{[IH]}}$.
The (\ref{PEQV}) means that reference price will slowly follow the trend as trading proceed.
This trend following stops only with $\Ket{\psi^{[IH]}}$ state switch,
what means a new spike in $I$ has been observed and this new spike
is now the ``most dramatic market observed''.

\begin{figure}[t]
  \includegraphics[width=16cm]{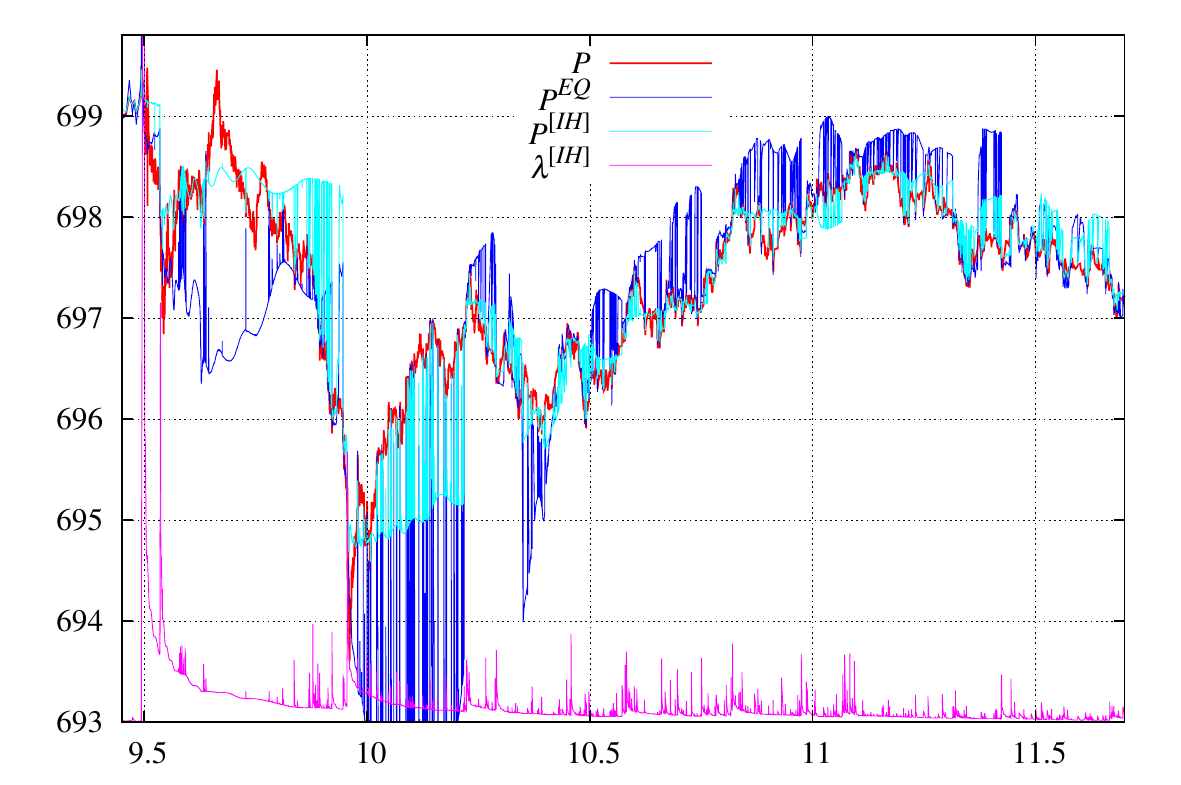}
  \caption{\label{PEQVFigure}
    Price $P$, price $P^{[IH]}$ (\ref{PIHGEV}),
    $P^{EQ}$ (\ref{PEQV})
    and maximal eigenvalue of (\ref{GEVdef})
    for
 AAPL stock on September, 20, 2012.
 The calculations in shifted Legendre basis with $n=12$ and $\tau$=256sec.
  The execution flow eigenvalue is scaled and shifted to $693$ to fit the chart.
  }
\end{figure}

An important feature of $P^{EQ}$
is that in 
(\ref{PEQV})
there are only the moments that
are calculated directly from sample using (\ref{dFSampleAll}):
$\Braket{Q_m I}$, $\Braket{Q_m pI}$, and $\Braket{Q_m V\frac{dp}{dt}}$.
This makes all the calculations easy.
In Fig. \ref{PEQVFigure} we present the 
$P^{EQ}$
along with  $P^{[IH]}$.
In \cite{2015arXiv151005510G}
the best directional indicator found was the difference
between last price and $P^{[IH]}$, without trending term.
As the price reaches some trading band --
it starts crossing $P^{[IH]}$ multiple times thus creating false signals.
The $P^{EQ}$ has a great advantage of extra trend following
contribution in (\ref{PEQV}), what very much suppresses false signals.
The result is also stable in situations when max $I$ ``switch'' is missed.

\subsection{\label{ExecutionFlowBasedDynamics}Execution Flow Driven Dynamics}

Consider ``exact differential'' action
with $-\Braket{\psi^{[IH]}\left|(p-P^{last})\frac{dV}{dt}\right|\psi^{[IH]}}$
term from (\ref{lagrangianV}):
\begin{align}
  {\cal S}^{ed}&=
  \Braket{\psi^{[IH]}\left|(P^{last}-p)I\right|\psi^{[IH]}}=
  \mathrm{Spur}\left\|\frac{d}{dt}(p-P^{last})I
  \middle|\rho_{JIH}\right\| \nonumber \\  
   &=\mathrm{Spur}\left\|I\frac{dp}{dt}\middle|\rho_{JIH}\right\|
  +\mathrm{Spur}\left\|p\frac{dI}{dt}\middle|\rho_{JIH}\right\|
  \label{SInodynamics}
\end{align}
Then ``actual'' action\footnote{
Note that if one put in (\ref{SI})
$\left\|\frac{d}{dt}(V^{last}-V) \frac{dp}{dt}\right\|$
and
$\left\|\frac{d}{dt}(P^{last}-p)I\right\|$
instead of $\left\|I\frac{dp}{dt}\right\|$ and
$\left\|p\frac{dI}{dt}\right\|$
the (\ref{SV}) is obtained.
} is considered as ``kinetic'' and ``potential'' terms split;
the kinetic term is defined as the one with \textsl{first derivative of price}
in $\mathrm{Spur}$ with $\left\|\rho_{JIH}\right\|$.
\begin{align}
  {\cal S}&=
  \mathrm{Spur}\left\|I\frac{dp}{dt}\middle|\rho_{JIH}\right\|
  -\mathrm{Spur}\left\|p\frac{dI}{dt}\middle|\rho_{JIH}\right\|
    \label{SI}
\end{align}
Here and below $\left\|\rho_{JIH}\right\|$ is the densitity matrix
corresponding to
the state ``since $\Ket{\psi^{[IH]}}$ till now'' (\ref{wfPartsdef}),
obtained from the polynomial $J\left({\psi^{[IH]}}^2\right)$ (\ref{JD1})
by applying
Theorem 3 of \cite{ArxivMalyshkinLebesgue},
see \texttt{\seqsplit{com/polytechnik/utils/BasisFunctionsMultipliable.java:getMomentsOfMeasureProducingPolynomialInKK\_MQQM}} for numerical implementation;
the $\left\|\rho_{JJIH}\right\|$ is the densitity matrix
corresponding to the polynomial $J\left(J\left({\psi^{[IH]}}^2\right)\right)$
(\ref{JD2});
\begin{align}
  \left\|\rho_{JIH}\right\|&=\left\|\rho_{J\left({\psi^{[IH]}}^2\right)}\right\| \label{rhoJIH} \\
  \left\|\rho_{JJIH}\right\|&=\left\|\rho_{J\left(J\left({\psi^{[IH]}}^2\right)\right)}\right\| \label{rhoJJIH}
\end{align}
see
\texttt{\seqsplit{com/polytechnik/freemoney/IandDM.java:\{QQDensityMatrix,QQDensityMatrix2\}}}
for numerical calculation of $\left\|\rho_{JIH}\right\|$
and $\left\|\rho_{JJIH}\right\|$
from $\psi^{[IH]}(x)$.
The second term in (\ref{SI}) does not depend on price shift $p\to p+const$
as with (\ref{iofuture}) boundary condition we have
$\mathrm{Spur}\left\|\frac{dI}{dt}\middle|\rho_{JIH}\right\|=0$.
The dynamics with (\ref{SI}) action is execution flow driven dynamics.
Let $I=\frac{dV}{dt}=const$ then the volume
is a linear functions on time $V-V^{last}= It$
and the price is constant $dp/dt=0$,
both terms in (\ref{SI}) are zero in any state;
moreover when $p(t)=I(t)$ the two terms are equal exactly for any $I(t)$.
This is different from the dynamics defined by (\ref{SV})
action where constant $I$ causes linear dependence of price on time.
With (\ref{SI}) action all changes in price 
are caused by changes in execution flow.
Our previous observations \cite{2015arXiv151005510G,2016arXiv160204423G}
show that
asset prices are
much more sensitive to execution flow $I$ (dynamic impact),
rather than to traded volume $V$ (regular impact).

Whereas the calculation of (\ref{SV}) action
was easy because the moments were calculated
directly from sample, the calculation of (\ref{SI})
is much more difficult. We can obtain directly from sample
only operator $\|pI\|$ and then, using (\ref{momsDefinitionByPartsPsiIBasis}),
operator $\left\|\frac{dpI}{dt}\right\|=\left\|I\frac{dp}{dt}\right\|
  +\left\|p\frac{dI}{dt}\right\|$:
\begin{align}
  \left\|I\frac{dp}{dt}\right\|
  -\left\|p\frac{dI}{dt}\right\|&=
  2\left\|I\frac{dp}{dt}\right\|-\left\|\frac{dpI}{dt}\right\|
  \label{pIdiff}
\end{align}
The problem left is to calculate the operator $\left\|I\frac{dp}{dt}\right\|$.
Currently we do not have a method to obtain it exactly.
The problem
is simplified by the fact that
we need not the operator $\left\|I\frac{dp}{dt}\right\|$ per se,
but just $\mathrm{Spur}\left\|I\frac{dp}{dt}\middle|\rho_{JIH}\right\|$,
what enables us to work with it's approximation.
See Appendix \ref{momentsCallculation} below for several approximations
for $\left\|I\frac{dp}{dt}\right\|$;
one can possibly try secondary sampling approach
of Appendix \ref{SecondarySampling}
as an alternative route.
Different approximations
give noticeably different results.
Nevetheless, assume we know the value of $\mathrm{Spur}\left\|I\frac{dp}{dt}\middle|\rho_{JIH}\right\|$ on past sample.
Then, assume one more observation with price $P^{EQ}$ is coming.
With the knowledge 
of future execution flow (\ref{iofuture}) we can put
 equal ``kinetic'' and ``potential'' terms in (\ref{SI}),
thus to obtain the value of price $P^{EQ}$ at which (\ref{SI}) is zero
(use (\ref{iofuture}) boundary condition
with (\ref{momsDefinitionByPartsPsiIBasis}) expansion):
\begin{align}
  \Delta_{I}&=\mathrm{Spur}\left\|I\frac{dp}{dt}\middle|\rho_{JIH}\right\|-\mathrm{Spur}\left\|p\frac{dI}{dt}\middle|\rho_{JIH}\right\| \nonumber \\
&=2\mathrm{Spur}\left\|I\frac{dp}{dt}\middle|\rho_{JIH}\right\|
    -\mathrm{Spur}\left\|\frac{dpI}{dt}\middle|\rho_{JIH}\right\|
  \label{PEQIDelta}\\
  P^{EQ}&=P^{last} 
  -\frac{\Delta_{I}}{\lambda^{[IH]}} \label{PEQI}\\
    &=2P^{last}-P^{[IH]}-\frac{2}{\lambda^{[IH]}}\mathrm{Spur}\left\|I\frac{dp}{dt}\middle|\rho_{JIH}\right\|
      \nonumber
\end{align}
In (\ref{PEQI}) all ``future'' price contributions are moved
to the left hand side and in the right hand side all the integration
is performed till last observed point with $p=P^{last}$.
Technically (\ref{PEQI}) means:
calculate the difference $\Delta_{I}$ (\ref{PEQIDelta})
on observed sample,
and if it is not zero -- the price
will move on $-\Delta_{I}/\lambda^{[IH]}$ to compensate.
Our experiments show that these two terms
are very close to each other and the value of $\Delta_{I}$ is small.
One may also try other states,
such as $\left\|\rho_{JJIH}\right\|$ (\ref{rhoJJIH}) to consider the operators in,
but the property of
operators
$\left\|I\frac{dp}{dt}\right\|$ and $\left\|p\frac{dI}{dt}\right\|$
being equal in some density matrix state seems to be special to the state
$\left\|\rho_{JIH}\right\|$, see 
Appendix \ref{doubleIntegration}
below for a study of the state $\left\|\rho_{JJIH}\right\|$.

\subsection{\label{Vdpdt}Local Volume Driven Dynamics}
Similarly to (\ref{SInodynamics})
one can
consider the term
$-\Braket{\psi^{[IH]}\left|(V-V^{last})\frac{dp}{dt}\right|\psi^{[IH]}}$
from (\ref{lagrangianV})
to be an ``exact differential'' action:
\begin{align}
  {\cal S}^{ed}&=
  \Braket{\psi^{[IH]}\left|(V^{last}-V)\frac{dp}{dt}\right|\psi^{[IH]}}=
  \mathrm{Spur}\left\|\frac{d}{dt}(V-V^{last})\frac{dp}{dt}
  \middle|\rho_{JIH}\right\| \nonumber \\  
   &=\mathrm{Spur}\left\|I\frac{dp}{dt}\middle|\rho_{JIH}\right\|
  +\mathrm{Spur}\left\|(V-V^{last})\frac{d^2p}{dt^2}\middle|\rho_{JIH}\right\|
  \label{SVdpnodynamics}
\end{align}
Then ``actual'' action is:
\begin{align}
  {\cal S}&=
   \mathrm{Spur}\left\|I\frac{dp}{dt}\middle|\rho_{JIH}\right\|
  -\mathrm{Spur}\left\|(V-V^{last})\frac{d^2p}{dt^2}\middle|\rho_{JIH}\right\|
  \label{SVLD}
\end{align}
The dynamics with (\ref{SVLD}) action is local volume driven dynamics.
Let $I=\frac{dV}{dt}=const$,
then the volume is a linear functions on time $V-V^{last}= It$.
Two terms equal
give quadratic dependence of price on time:
$P=\alpha t^2+C$;
then two terms in (\ref{SVLD}) are equal
exactly in any state:
$I\frac{dp}{dt}=(V-V^{last})\frac{d^2p}{dt^2}$.
There is a similar problem with calculation of the derivatives 
to the one considered above. Taking into account
$\left\|\frac{d}{dt}(V-V^{last})\frac{dp}{dt}\right\|=
\left\|I\frac{dp}{dt}\right\|+\left\|(V-V^{last})\frac{dp^2}{dt^2}\right\|$ obtain:
\begin{align}
\left\|I\frac{dp}{dt}\right\|
-\left\|(V-V^{last})\frac{d^2p}{dt^2}\right\|
&=2\left\|I\frac{dp}{dt}\right\|-\left\|\frac{d}{dt}(V-V^{last})\frac{dp}{dt}\right\|
\label{pvexpansiondiff}
\end{align}
As with (\ref{pIdiff}) above
the second term is obtained by applying (\ref{momsDefinitionByPartsPsiIBasis})
to the moments $\left\|(V-V^{last})\frac{dp}{dt}\right\|$
that are available directly from sample.
Consider one more observation with price $P^{EQ}$ coming.
With the knowledge of future execution flow (\ref{iofuture})
we can obtain the equilibrium price:
\begin{align}
  \Delta_{V}&=\mathrm{Spur}\left\|I\frac{dp}{dt}\middle|\rho_{JIH}\right\|
  -\mathrm{Spur}\left\|(V-V^{last})\frac{d^2p}{dt^2}\middle|\rho_{JIH}\right\|
  \nonumber \\
  &=2\mathrm{Spur}\left\|I\frac{dp}{dt}\middle|\rho_{JIH}\right\|
    -\mathrm{Spur}\left\|\frac{d}{dt}(V-V^{last})\frac{dp}{dt}\middle|\rho_{JIH}\right\|  
  \label{PEQLocalVolumeDelta}\\
  P^{EQ}&=P^{last}
  -\frac{\Delta_{V}}{\lambda^{[IH]}} \label{PEQLocalVolume}
\end{align}
Technically (\ref{PEQLocalVolume}) means:
calculate the difference $\Delta_{V}$ (\ref{PEQLocalVolumeDelta})
on observed sample,
and if it is not zero -- the price
will move on $-\Delta_{V}/\lambda^{[IH]}$ to compensate.

\subsection{\label{dVPTotal}Total Lagrangian  Driven Dynamics}
The same as in Sections \ref{ExecutionFlowBasedDynamics} and \ref{Vdpdt} above
we can consider the entire $L(p,V)$ from (\ref{lagrangianV})
in $\Ket{\psi^{[IH]}}$ state $-\Braket{\psi^{[IH]}|L(p,V)|\psi^{[IH]}}$
to be an ``exact differential'' action:
\begin{align}
  {\cal S}^{ed}&=
  \Braket{\psi^{[IH]}\left|(P^{last}-p)I\right|\psi^{[IH]}}+
  \Braket{\psi^{[IH]}\left|(V^{last}-V)\frac{dp}{dt}\right|\psi^{[IH]}} \nonumber \\
  &=2\mathrm{Spur}\left\|I\frac{dp}{dt}\middle|\rho_{JIH}\right\|
  +\mathrm{Spur}\left\|p\frac{dI}{dt}\middle|\rho_{JIH}\right\|
  +\mathrm{Spur}\left\|(V-V^{last})\frac{d^2p}{dt^2}\middle|\rho_{JIH}\right\|
  \label{STotalnodynamics}
\end{align}
This is actually the same expression
as (\ref{SVnodynamics}) above (with changed sign),
but split differently into ``kinetic'' and ``potential'' energy terms.
The ``actual'' action
is then obtained by changing the sign of
``potential'' contributions:
\begin{align}
  {\cal S}&=
 2\mathrm{Spur}\left\|I\frac{dp}{dt}\middle|\rho_{JIH}\right\|
 -\left\lgroup
   \mathrm{Spur}\left\|p\frac{dI}{dt}\middle|\rho_{JIH}\right\|
   +\mathrm{Spur}\left\|(V-V^{last})\frac{d^2p}{dt^2}\middle|\rho_{JIH}\right\|
 \right\rgroup
 \label{STotal}
\end{align}
The dynamics for $I=\frac{dV}{dt}=const$ can be obtained in a regular way.
The volume is a linear functions on time $V-V^{last}= It$.
The price with ``kinetic'' and ``potential'' terms equal
gives cubic dependence of price on time:
$P=\alpha t^3+C$;
the terms in (\ref{STotal}) are equal
exactly in any state:
$2I\frac{dp}{dt}=0+(V-V^{last})\frac{d^2p}{dt^2}$.

The $\Delta_{T}$ correponds to the sum of (\ref{PEQIDelta}) and (\ref{PEQLocalVolumeDelta}):
\begin{align}
  \Delta_{T}&=\Delta_{I}+\Delta_{V} \label{PEQTotalDelta} \\
  &=4\mathrm{Spur}\left\|I\frac{dp}{dt}\middle|\rho_{JIH}\right\|
  -\mathrm{Spur}\left\|\frac{dpI}{dt}\middle|\rho_{JIH}\right\|
  -\mathrm{Spur}\left\|\frac{d}{dt}(V-V^{last})\frac{dp}{dt}\middle|\rho_{JIH}\right\|
  \label{LAllTermsDelta}  \\
   P^{EQ}&=P^{last} 
   -\frac{\Delta_{T}}{\lambda^{[IH]}} \label{PEQTotal} 
\end{align}
Whereas this ``Total Lagrangian Driven''
and ``Volume Driven'' dynamics of Section \ref{VolumeBasedDynamics} above
use the same ``exact differential action'':
(\ref{STotalnodynamics})
and
(\ref{SVnodynamics}),
they generate different dynamics
as we differently split action into ``kinetic'' and ``potential'' terms.
The  dynamics of Section \ref{VolumeBasedDynamics}
is of trend following type, the direction changes
only when $\Ket{\psi^{[IH]}}$ ``switches''.
The direction of ``Total Lagrangian Driven'' dynamics takes
into account a number of factors in (\ref{STotal}),
thus it may reverse the direction
even without a ``switch'' in $\Ket{\psi^{[IH]}}$.

\section{\label{DynamicsSelection}On Selection Of A Type Of The Dynamics }
In Sections \ref{VolumeBasedDynamics}, \ref{ExecutionFlowBasedDynamics},
\ref{Vdpdt}, and \ref{dVPTotal}
we chose an ``exact differential'' action from
which price dynamics was determined.
The ``volume driven'' dynaimcs of Section \ref{VolumeBasedDynamics}
stays separately as it is a trend--following model with
automatic time--scale selection of Section \ref{TimeScale},
it is the simplest ``practical'' model.
An important feature of possible ``action'' of the forms:
(\ref{SI}), (\ref{SVLD}), and (\ref{STotal})
is that all of them include fluctuations of execution flow $dI/dt$.
Now we have to choose which one corresponds to market dynamics
most closely.
Technically, we have three calculated characteristics:
$\mathrm{Spur}\left\|\frac{dpI}{dt}\middle|\rho_{JIH}\right\|$
and
$\mathrm{Spur}\left\|\frac{d}{dt}(V-V^{last})\frac{dp}{dt}\middle|\rho_{JIH}\right\|$
that are obtained exactly and
$\mathrm{Spur}\left\|I\frac{dp}{dt}\middle|\rho_{JIH}\right\|$
that is obtain from an approximation, such as in the
Appendix \ref{momentsCallculation}.
Now we need to select from them the most appropriate linear
combination to obtain the
future price
according to (\ref{PEQIDelta}), (\ref{PEQLocalVolumeDelta}),
or (\ref{PEQTotalDelta}).

\begin{figure}
  \includegraphics[width=16cm]{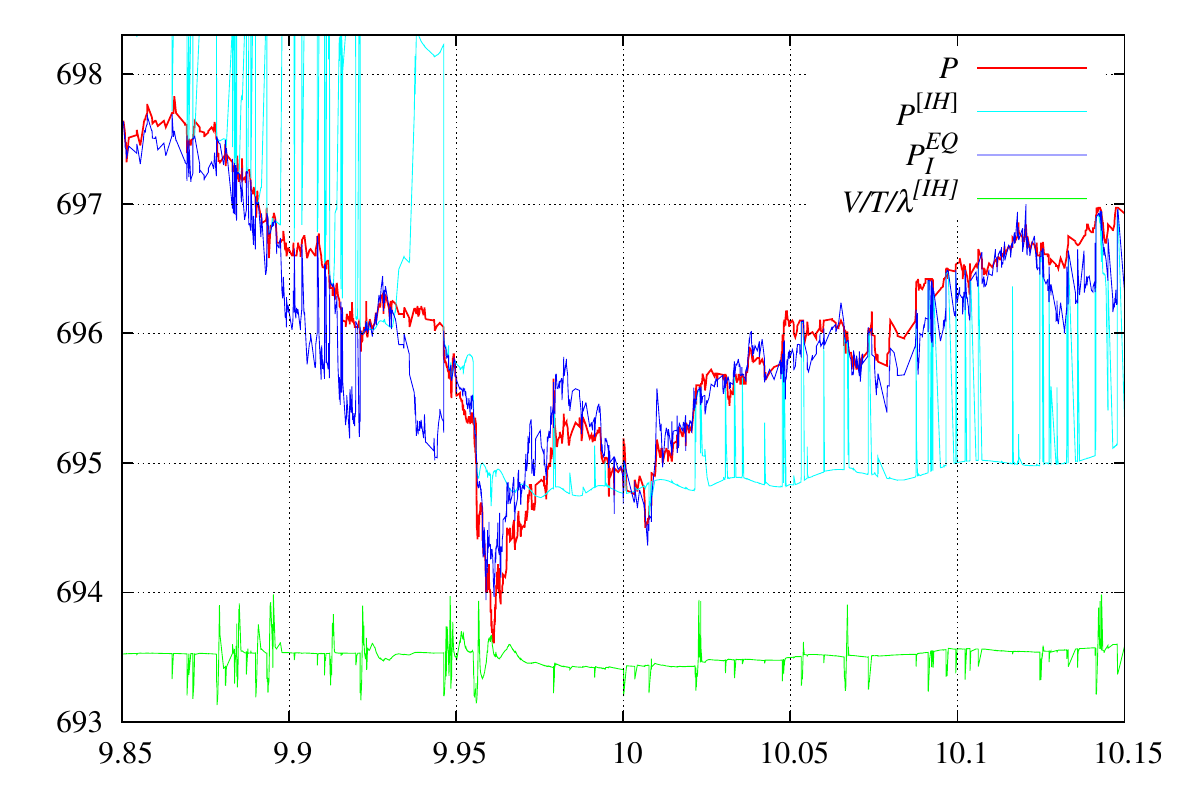}
  \caption{\label{PEQIPlot}
    Price $P$, price $P^{[IH]}$ (\ref{PIHGEV}),
    and $P^{EQ}$ (\ref{PEQIDelta})
    for
    AAPL stock on September, 20, 2012.
    The ratio of aggregated $\frac{V}{T}$ execution flow and
    $\lambda^{[IH]}$ is also presented (shifted up to 693); it is about $1/2$ in $\Ket{\psi^{[IH]}}$
    state.
 The calculations in shifted Legendre basis with $n=12$ and $\tau$=256sec.
  }
\end{figure}

Experiments show that only execution flow driven dynamics
of
(\ref{SI}) form
can possibly work to determine the future price.
This type of dynamics
has only two contributing terms:  $I\frac{dp}{dt}$ and $p\frac{dI}{dt}$,
that are very close to each other
in $\|\rho_{JIH}\|$ state.
The sum of them is equal exactly
to $\lambda^{[IH]}(P^{last}-P^{[IH]})=\mathrm{Spur}\left\|\frac{dpI}{dt}\middle|\rho_{JIH}\right\|$ (the result of \cite{2015arXiv151005510G}),
and their difference
determines (\ref{PEQIDelta}) future price (a new result of this paper).
In Fig. (\ref{PEQIPlot}) a demonstration
of $P^{EQ}$ (\ref{PEQIDelta}) (field
\texttt{\seqsplit{com/polytechnik/freemoney/PFuture.java:PEQ\_I\_fromDPI}},
obtained with
$\|I\frac{dp}{dt}\|$ from (\ref{DPwithI}) is presented.
The result is almost identical to
(\ref{momsDefinitionByPartsPDivIMultiplyI2BasisModifPractical}) approximation,
field \texttt{\seqsplit{com/polytechnik/freemoney/PFuture.java:PEQ\_I\_fromI2DtpDivI}},
and the $\Delta_{I}$ is slightly lower with (\ref{DPwithITwoMultiplies})).
This is the only ``advancing''
(not lagging!) indicator we managed to obtain so far.
A very important feature is that
operators 
$\left\|p\frac{dI}{dt}\right\|$
and
$\left\|I\frac{dp}{dt}\right\|$
from (\ref{pIdiff})
are very close in  $\rho_{JIH}$ state,
and the difference determines future direction (\ref{PEQIDelta}).
This corresponds to
$\frac{2}{\lambda^{[IH]}}\mathrm{Spur}\left\|I\frac{dp}{dt}\middle|\rho_{JIH}\right\|\approx \mathrm{Spur}\left\|\frac{dp}{dt}\middle|\rho_{JIH}\right\|$;
the ratio of aggregated
execution flow $\frac{V}{T}=\frac{\mathrm{Spur}\left\|I\middle|\rho_{JIH}\right\|}{\mathrm{Spur}\left\|\rho_{JIH}\right\|}$ and execution flow $\lambda^{[IH]}$ is about
$0.5$ in $\Ket{\psi^{[IH]}}$ state, see green line in Fig. \ref{PEQIPlot}.

One may also consider other states, 
e.g. from \cite{MalMuseScalp}: ``Appendix C: The state of maximal aggregated execution flow $V/T$'',
that corresponds to eigenvalue problem
\begin{align}
  \Ket{V\middle|\psi^{[i]}}=\lambda^{[i]}\Ket{T\middle|\psi^{[i]}} \label{GEVVTdef}
\end{align}
and using this $V/T$ maximal $\lambda$ in boundary condition (\ref{iofuture}).
A remarkable feature of this state
$\Ket{\psi^{[VT]}}$
is that in it $dV/dT=V/T=\lambda^{[V/T]}$:
\begin{align}
 \frac{\Braket{\psi^{[VT]}\left|I\right|\psi^{[VT]}}}{
    \Braket{\psi^{[VT]}|\psi^{[VT]}}}
    &=\frac{\Braket{\psi^{[VT]}\left|V\right|\psi^{[VT]}}}{
    \Braket{\psi^{[VT]}\left|T\right|\psi^{[VT]}}    
 }=
  \frac{\mathrm{Spur}\left\|I\middle|\rho_{JVT}\right\|}{\mathrm{Spur}\left\|\rho_{JVT}\right\|}=\lambda^{[V/T]}\le\lambda^{[IH]}
  \label{VTequalsI}
\end{align}
what allows us to consider ``double integrated'' type of
the density matrix we considered in Appendix \ref{doubleIntegration}.
In this $\Ket{\psi^{[VT]}}$ state we typically have
$\mathrm{Spur}\left\|I\frac{dp}{dt}\middle|\rho_{JVT}\right\|\approx \mathrm{Spur}\left\|\frac{dpI}{dt}\middle|\rho_{JVT}\right\|$, no ``advancing''
information we managed to obtain so far.

When asked about ``direct application'' of the solution presented
in Fig. \ref{PEQIPlot},
an ``advancing'' indicator,
to practical trading --
it is not ready for two reasons:
\begin{itemize}
\item The result strongly depends on approximation used
  for $\left\|I\frac{dp}{dt}\right\|$,
  see Appendix \ref{momentsCallculation}.
\item In Fig. \ref{PEQIPlot} we have shown only the interval,
  where predicted future price is substantially different from last price.
  For large intervals of this trading session
  the difference between predicted and last price
  is very small,
  this means the relation
  $\mathrm{Spur}\left\|p\frac{dI}{dt}\middle|\rho_{JIH}\right\|=
  \mathrm{Spur}\left\|I\frac{dp}{dt}\middle|\rho_{JIH}\right\|$
  holds almost exactly in $\Ket{\psi^{[IH]}}$ state.
\end{itemize}
This makes us to conclude that execution flow driven
dynamics of Section \ref{ExecutionFlowBasedDynamics},
while is a very promising one and is the only one
we obtained, that
can possibly provide an ``advancing'' indicator -- it still
requires more work for practical applications.
These two directions are the most promising:
\begin{enumerate}
\item Select other eigenvalue problem (\ref{GEVgeneralMatrix})
  that will be determining the time scale.
  The best what we found is (\ref{GEVdef}) with the state $\Ket{\psi^{[IH]}}$
  of maximal eigenvalue $\lambda^{[IH]}$.
  All other tried have been worse.
\item Select directional indicator.
  The best indicator we found so far is the (\ref{PEQIDelta}).
\end{enumerate}

\section{\label{DirectionInfo}On Practical Source Of Directional Information}
Whereas our attempts above to obtain an ``advancing'' indicator
were promising but not ready yet for practical trading,
a lagging directional indicator with automatic
time--scale selection of Section \ref{VolumeBasedDynamics}, is ready
and can be applied to trading.
This indicator (\ref{PEQV}) is
the $P^{[IH]}$ price plus trend-following factor
that is proportional to $dp/dt$ in $\Ket{\psi^{[IH]}}$ state:
\begin{align}
  P^{EQ}&=P^{[IH]}-  \frac{1}{\lambda^{[IH]}}
  \Braket{\psi^{[IH]}\left|(V-V^{last})\frac{dp}{dt}\right|\psi^{[IH]}}
  \label{PEQVFormula}
\end{align}
The formula automatically selects the time scale
from the interval $[\tau/n:\tau]$ (in Shifted Legendre basis)
or from the interval $[\tau:n\tau]$ (in Laguerre basis).
The difference between last price and $P^{EQ}$ determines the trend.
Trend ``switch'' occurs instantly as a ``switch'' in $\Ket{\psi^{[IH]}}$
of (\ref{GEVdef}) eigenproblem.
In Fig. \ref{PVforVsurrogate} below the (\ref{PEQVFormula})
is presented in higher resolution than in Fig. \ref{PEQVFigure} above.
The situation when $P^{EQ}$ is close to $P^{last}$ corresponds
to no information about the future (\ref{dIeq0Psi}) situation.
Typically all the directional signals should be ignored (\ref{dIeq0Psi}) when
\begin{align}
  \Braket{\psi_0|\psi^{[IH]}}^2&\gtrsim 0.1
    \label{SignalsToIgnore}
\end{align}
as this corresponds to little information about the future available.

\section{\label{HowItWorks}A Brief Description Of The Algorithm}
Whereas a theory presented is this work may look rather
complicated, it's computer implementation is very straightforward.
It is way simpler than multiple systems of other people
the authors have seen in diversity
$\stackrel{\cdot\cdot}{\frown}$.
Technically an implementation of
the theory requires an integration to calculate the
moments from timeserie sample (\ref{dFSampleAll}),
polynomials multiplication (\ref{multiplicationOperator})
to calculate the matrices from moments,
and solving an eigenproblem (\ref{GEVdef}) for time scale.
There is no magic,
simple and precise description of an algorithm
implementing the theory is this:
On each (Time, Execution Price, Shares Traded) tick coming do the following:
\begin{enumerate}
\item Have an integrator that
  on each tick coming recurrently updates an internal states to calculate the moments:
  $\Braket{Q_m}$, $\Braket{Q_mI}$, $\Braket{Q_mpI}$,
  and
  $\Braket{Q_m V\frac{dp}{dt}}$;
  see \texttt{\seqsplit{com/polytechnik/freemoney/CommonlyUsedMoments.java}}.
\item Using these moments construct the matrices
  $\Braket{Q_j|Q_k}$, $\Braket{Q_j|I|Q_k}$, $\Braket{Q_j|pI|Q_k}$,
  and $\Braket{Q_m \frac{d}{dt}V\frac{dp}{dt}}$
  by applying
  multiplication operator (\ref{multiplicationOperator}),
  then solve generalized eigenvalue problem (\ref{GEV})
  to find the state of maximal execution flow $\Ket{\psi^{[IH]}}$
  determining the time scale,
  see \texttt{\seqsplit{com/polytechnik/freemoney/FreeMoneyForAll.java}}.
\item  Construct a polynomial
  corresponding to the state ``since $\Ket{\psi^{[IH]}}$ till now'' (\ref{wfPartsdef}),
  then obtain corresponding density matrix $\|\rho_{J\left({\psi^{[IH]}}^2\right)}\|$
  using
  Theorem 3 of \cite{ArxivMalyshkinLebesgue},
  see \texttt{\seqsplit{com/polytechnik/freemoney/IandDM.java}}.

\item Obtain (\ref{PEQVFormula}) price as directional indicator.
  Use (\ref{SignalsToIgnore}) as non-applicability criterion.
  See
  \texttt{\seqsplit{com/polytechnik/freemoney/PFuture.java:PEQV\_from\_M}}.

\item Optionally.
  Try (\ref{PEQI}), but this calculation requires an approximation
  for $\left\|I\frac{dp}{dt}\right\|$
  and is much more sensitive to time-scale selection than (\ref{PEQVFormula}).
  The result for ``advancing'' directional indicator (\ref{PEQI})
  is not always satisfactory.

\end{enumerate}
See Appendix \ref{SoftwareDescription} below with a description of the
\href{http://www.ioffe.ru/LNEPS/malyshkin/AMuseOfCashFlowAndLiquidityDeficit.zip}{software}
implementing the algorithm.
This software reads a sequence of (Time, Execution Price, Shares Traded)
ticks (line after line, one tick per line),
and for every tick read
prints the results.

\section{\label{conclusion}Conclusion}
An approach to obtain directional information
from a sequence of past transactions
with an automatic time--scale selection
from execution flow $I=dV/dt$
is presented.
Whereas a regular moving average has a built-in fixed time scale,
the approach of this paper uses the state of maximal execution flow (\ref{GEVdef})
to automatically determine the one.
Contrary to regular moving average
the developed approach has $n-1$ internal degrees of freedom
to adjust averaging weight according to spikes in
 execution flow $I=dV/dt$.
These internal degrees of freedom allow
to obtain
an immediate ``switch'', what is not possible in regular moving average
that always has a $\tau$-proportional time delay, lagging indicator.
For a problem of dimension $n$
in Shifted Legendre basis the system automatically selects
the time scale from the interval $[\tau/n:\tau]$,
and in Laguerre basis from the interval $[\tau:n\tau]$.
Among unsolved problem we would note a selection
of optimal $\left\|I\frac{dp}{dt}\right\|$ interpolation
to obtain an advancing price from (\ref{PEQIDelta}),
see Appendix \ref{momentsCallculation},
and studying a possibility to ``split'' some average value based on some other
operator spectrum, see Appendix \ref{dIprojections}, Eq. (\ref{splitProj}).
The software implementing the theory
\href{http://www.ioffe.ru/LNEPS/malyshkin/AMuseOfCashFlowAndLiquidityDeficit.zip}{is available}
from the authors.
Among directly applicable to trading results we would note 
the price (\ref{PEQVFormula})
that includes both ``switching'' and ``tending'' contributions,
see Fig. \ref{PVforVsurrogate}.

A generalization of the developed theory to a multi asset universe
creates a number of new opportunities.
Now from a sequence of past transactions $l=1\dots M$ for $N_a$
financial instruments:
\begin{align}
  &\left(t_l,p^{(a)}_l,dV^{(a)}_l\right) & a=1\dots N_a
  \label{seqMultiStock}
\end{align}
one can construct $N_a$ execution
flow operators $\Braket{Q_j\left|I^{(a)}\right|Q_k}$,
each one with it's own state $\Ket{{\psi^{[IH]}}^{(a)}}$ of maximal $I^{(a)}$
and corresponding to it $\rho_{JIH}^{(a)}$.
In addition to this an operator of capital-flow index 
\begin{align}
  \widetilde{I}&=
  \sum\limits_{a=1}^{N_a} p^{(a)}\frac{dV^{(a)}}{dt}
  =\sum\limits_{a=1}^{N_a} p^{(a)}I^{(a)}
  \label{seqMultiStockIndex}
\end{align}
can be constructed to determine market overall activity;
it also has it's own state $\Ket{{\psi^{[IH]}}^{\widetilde{I}}}$
of maximal $\widetilde{I}$\footnote{
A one of self-evident trading strategies:
when current value of $\widetilde{I}$ is large
$\Braket{\psi_0|{\psi^{[IH]}}^{\widetilde{I}}}^2\gtrsim 0.8$
select the assets $a$ with currently low execution flow 
$\Braket{\psi_0|{\psi^{[IL]}}^{(a)}}^2\gtrsim 0.5$
as ``lagging'' and soon to follow in the direction of the market.
}.
Critically important that all these $\Ket{\psi}$ are in the same basis
(the one of (\ref{Xbasis})) and their scalar products $\Braket{\psi|\phi}$
can be readily calculated. Technically this means
we can independently use $N_a$ integrators 
\texttt{\seqsplit{com/polytechnik/freemoney/CommonlyUsedMoments.java}},
where each one calculates the moments $\Braket{Q_mf^{(a)}}$ of it's
own single asset $a=1\dots N_a$,
and then, from here, all the
cross-asset characteristics
can be calculated via projections!
For example: how similar is the state of high execution flow of asset $a$
and the
one of asset $b$ ?---
it is just a regular scalar product
of two wavefunctions
$\Braket{{\psi^{[IH]}}^{(a)}|{\psi^{[IH]}}^{(b)}}^2$;
``correlated'' assets are not the assets which prices ``go together''
but the assets with simultaneous spikes in execution flow.
In addition to simultaneously criterion (projection)
a criterion for ``which one came earlier:
a spike in $\|I^{(a)}\|$ or a spike in $\|I^{(b)}\|$''
can be written in a similar way:
$\frac{\Braket{{\psi^{[IH]}}^{(a)}|(t_{now}-t)I^{(a)}|{\psi^{[IH]}}^{(a)}}}
{\Braket{{\psi^{[IH]}}^{(a)}|I^{(a)}|{\psi^{[IH]}}^{(a)}}}
-
\frac{\Braket{{\psi^{[IH]}}^{(b)}|(t_{now}-t)I^{(b)}|{\psi^{[IH]}}^{(b)}}}
{\Braket{{\psi^{[IH]}}^{(b)}|I^{(b)}|{\psi^{[IH]}}^{(b)}}}$
obtained from directly sampled
moments $\Braket{Q_m(t_{now}-t)I^{(a)}}$
and
$\Braket{Q_m(t_{now}-t)I^{(b)}}$,
or as
$\mathrm{Spur}\|\rho^{(a)}_{JIH}\|-\mathrm{Spur}\|\rho^{(b)}_{JIH}\|$
what does not require other moments.
There are several alternative forms of ``distance'' to determine
which $\Ket{\psi}$ happened earlier,
see \cite{ArxivMalyshkinMuse}, ``Appendix A: Time--Distance Between $\Ket{\psi}$ States''.

In a multi asset univers complexity of calculations
growths \textsl{linearly} with $N_a$, hence the value
of $N_a$ can be very high even for realtime processing.
Moreover, as every integrator
\texttt{\seqsplit{com/polytechnik/freemoney/CommonlyUsedMoments.java}}
works independently, the problem can be easily parallelized
to run each integrator on a separate core.
Then all the cross-asset characteristics can be obtained
from individual asset data
(the moments from \texttt{\seqsplit{com/polytechnik/freemoney/CommonlyUsedMoments.java}} instance)
with standard linear algebra operations such as
projection (scalar product), taking the difference
between two $\mathrm{Spur}\|\rho_{JIH}\|$ to determine the distance,
or considering
some other operator (e.g. capital-flow index (\ref{seqMultiStockIndex}))
in a state like $\Ket{\psi}$,  $\|\rho_{J(\psi^2)}\|$,
or $\|\rho_{J(J(\psi^2))}\|$.

We see an application of this paper theory to multi asset universe
as the most promising direction of future research.
The simplest, but really good, indicator is
the ${P^{last}}^{(a)}-{P^{EQ}}^{(a)}$ indicator (\ref{PEQVFormula})
calculated for each $a=1\dots N_a$ asset then all summed
with the ${\lambda^{[IH]}}^{(a)}$ weights
for the terms in the sum to have the dimension of capital flow.

\begin{acknowledgments}
  This research was supported by
  Autretech group\cite{AutretechGroup}, \href{https://xn--80akau1alc.xn--p1ai/}{www.атретек.рф}.
  We thank our colleagues from Autretech R\&D department
  who provided insight and expertise that greatly assisted the research.
  Our grateful thanks are also extended
  to Mr. Gennady Belov for his methodological support in doing the data analysis.
\end{acknowledgments}

\appendix
\section{\label{momentsCallculation}On Calculation of $\Braket{Q_m I\frac{dp}{dt}}$
  moments from $\Braket{Q_m pI}$ sampled moments.}
A theory developed in this paper works primarily with
an observable $f$
and corresponding
operator (matrix)
$\Braket{Q_j|f|Q_k}$, $j,k=0\dots n-1$ that is obtained by applying
multiplication operator:
\begin{align}
  Q_j Q_k&=\sum_{m=0}^{j+k}c_m^{jk}Q_m
  \label{multiplicationOperator}
\end{align}
to sampled moments $\Braket{Q_mf}$, $m=0\dots 2n-2$.
The moments are defined with $Q_m(x)$ being a polynomial of order $m$
and integration measure $\omega(t)\,dt$ having
the support $t\in [-\infty\dots t_{now}]$:
\begin{align}
  \Braket{Q_m f}&=\int\limits_{-\infty}^{t_{now}} dt\,\omega(t)Q_m(x(t)) f(t)
  \label{momsDefinition}
\end{align}
In this paper we use:
$\omega(t)$ is decaying exponent
and
$x(t)$ is either linear or exponential function on time:
\begin{align}
  \omega(t)&=\exp\left(-(t_{now}-t)/\tau\right)
  \label{Wbasis} \\
  x(t)&=
  \begin{cases}
    (t-t_{now})/\tau & \text{Laguerre basis} \\
    \exp\left(-(t_{now}-t)/\tau\right) & \text{shifted Legendre basis}
  \end{cases}
  \label{Xbasis}
\end{align}
These two bases correspond to a more general
(also analytically approachable)
form
$x(t)=\exp\left(-(t_{now}-t)/\tau^*\right)$,
where $\omega(t)$ and $x(t)$ are both
exponential functions on $t$ but with different time scales:
$\tau$ and $\tau^*$. Other forms can also be considered.

If we want to consider $df/dt$ moments, then put it to
(\ref{momsDefinition}) and do an integration by parts:
\begin{align}
  \Braket{Q_m \frac{df}{dt}}&=
f(t_{now})\omega(x_0)Q_m(x_0)-
  \int\limits_{-\infty}^{t_{now}} dt\,f(t)\frac{d}{dt}\omega(x(t))Q_m(x(t))
  \label{momsDefinitionByParts}
\end{align}
if $\frac{d}{dt}\omega(x(t))Q_m(x(t))$
is equal to the same weight multiplied by a polynomial: $\omega(x)P(x)$ then
 the moments of $df/dt$ can be obtained from the
 moments of $f$
 according to (\ref{momsDefinitionByParts}).
The key element is
an existence of $\mathrm{ED}(\cdot)$,
a polynomial--to--polynomial mapping function
(it is obtained as a derivative of a polynomial multiplied by the weight):
\begin{align}
\frac{d}{dt}\omega(x(t))\psi(x(t))\varphi(x(t))&=
\omega(x)\left[
  \mathrm{ED}(\psi)\varphi+
  \psi\mathrm{ED}(\varphi)\right]
\label{ddtPsi2} \\
  \mathrm{ED}(\psi(x))&=
  \begin{cases}
    \displaystyle
    \frac{d\psi(x)}{dx}+\frac{1}{2}\psi(x)
    &\text{Laguerre basis} \\[1em]
    \displaystyle
    x\frac{d\psi(x)}{dx}+\frac{1}{2}\psi(x)
    &\text{shifted Legendre basis}
  \end{cases}
  \label{EDforDifferentBases}
\end{align}
where the time-derivative of a polynomial
multiplied by a weight
is represented by the \textsl{same weight}
multiplied by other polynomial.
The (\ref{momsDefinitionByParts})
corresponds to $\psi=1$ and $\varphi=Q_m$.

For the two bases we consider in this paper
it is also possible to obtain $\Braket{Q_m f}$ moments
from $\Braket{Q_m \frac{df}{dt}}$ moments using integration by parts,
see \cite{MalMuseScalp}, section ``Basic Mathematics'',
about $J(\cdot)$ polynomial-to-polynomial mapping such that
for an arbitrary polynomial $P(x)$:
\begin{align}
  \int\limits_{-\infty}^{t}P(x(t^{\prime}))\omega(t^{\prime})dt^{\prime}
  =\omega(t)J(P) \label{wfPartsdef}
\end{align}
For the bases we use such a polynomial-to-polynomial transform exists:
\begin{align}
  J(P)&=
  \begin{cases}
    {\displaystyle \frac{1}{\exp(x)}\int\limits_{-\infty}^x P(x^{\prime}) \exp(x^{\prime}) dx^{\prime}}
    &\text{Laguerre basis} \\
    {\displaystyle \frac{1}{x}\int\limits_0^x P(x^{\prime})dx^{\prime}}
    &\text{shifted Legendre basis}
  \end{cases}
  \label{JPDforDifferentBases}
\end{align}
A remarkable feature of this transform
is that since $\Braket{(f(t_{now})-f)P}=\Braket{J(P)\frac{df}{dt}}$
and $J(P)$ is also a polynomial, thus an average with it can be converted
to a density matrix average,
any average
$\Braket{P f}$ can be represented as
the spur from a product of operator $\|df/dt\|$
and a density matrix $\|\rho\|$:
\begin{align}
  \Braket{P f}&=\mathrm{Spur} \left\|\rho\middle|\frac{df}{dt}\right\|
  \label{spurFromP}
\end{align}
This way any average of
an observable $f$ can be calculated
as operator $\|df/dt\|$ averaged in some
\href{https://en.wikipedia.org/wiki/Quantum_state#Mixed_states}{mixed state} $\rho$
obtained from the polynomial $P(x)$.
Note, that $J(\cdot)$ transform can be applied in chain:
\begin{align}
  \int\limits_{-\infty}^{t_{now}}\frac{df}{dt}J(P)\omega(t)dt&=
  f\Big|_{x_0}J(P)\Big|_{x_0}\omega(x_0) 
  -\int\limits_{-\infty}^{t_{now}}f P(x(t))\omega(t)dt
  \label{JD1} \\
  \int\limits_{-\infty}^{t_{now}}\frac{d^2f}{dt^2}J(J(P))\omega(t)dt&=
  \frac{df}{dt}\Big|_{x_0}J(J(P))\Big|_{x_0}\omega(x_0) 
  -f\Big|_{x_0}J(P)\Big|_{x_0}\omega(x_0) 
  +\int\limits_{-\infty}^{t_{now}}f P(x(t))\omega(t)dt
  \label{JD2}
\end{align}

The moments of  $f$ are usually obtained
from direct sampling of all available observations $l=1\dots M$
in a timeserie:
\begin{align}
  \Braket{Q_m f}&=\sum_{l=1}^{M} f(t_l) Q_m(x(t_l)) \omega(t_l)
  \left[t_l-t_{l-1}\right]
\label{FSampleAll}
\end{align}
the moments of a derivative  $df/dt$
can also be obtained from direct sampling:
\begin{align}
  \Braket{Q_m \frac{df}{dt}}&=\sum_{l=1}^{M}
   Q_m(x(t_l)) \omega(t_l) \left[f(t_l)-f(t_{l-1})\right]
\label{dFSampleAll}
\end{align}
See \cite{ArxivMalyshkinMuse}, section ``Basis Selection'', for
one more basis: price basis: $Q_k(t)=p^k(t)$. It has no $\mathrm{ED}(\cdot)$ and $J(\cdot)$
operators available, but has similar sampling formula.
Given a good choice of basis polynomials:
\begin{align}
  Q_m(x)&=
  \begin{cases}
    L_m(-x) & \text{Laguerre basis} \\
    P_m(2x-1) & \text{shifted Legendre basis}
  \end{cases}
  \label{basesPol}
\end{align}
one can calculate (with double precision arithmetic) the moments to a very high order $m\lesssim 50$ (limited by the divergence of $c_m^{jk}$ multiplication coefficients
(\ref{multiplicationOperator}))
in Laguerre basis, and $m\lesssim 150$ (limited by poorly conditioned matrices)
in shifted Legendre basis;
Chebyshev polynomials $T_m(2x-1)$ also provide very stable calculations in shifted Legendre basis
(Chebyshev polynomials have perfectly stable multiplication:
all $c_m^{jk}=0$ except $c_{j-k}^{jk}=c_{j+k}^{jk}=0.5$, $j\ge k$).
The result is invariant with respect
to basis choice,  $Q_m(x)=x^m$ and the ones from (\ref{basesPol})
give \textsl{identical} results, but numerical stability
can be drastically different\cite{beckermann1996numerical,2015arXiv151005510G}.

Moments calculated from market data timeserie using Eqs. (\ref{FSampleAll}) and (\ref{dFSampleAll})
are the cornerstone of our theory. The most important are the
moments of execution flow $I=dV/dt$, they are obtained from (\ref{dFSampleAll})
by putting the volume as $f=V$,
thus the moments $\Braket{Q_m I}$
are obtained from
timeserie sample;
the matrix $\Braket{Q_j|I|Q_k}$
is obtained from them using
multiplication operator (\ref{multiplicationOperator}).
The matrix $\Braket{Q_j|Q_k}$ is known analytically.
These two matrices are volume- and time- averaged products of two basis functions.
A generalized eigenvalue problem (\ref{GEVgeneral}) is then formulated:
\begin{align}
\Ket{I\middle|\psi^{[i]}}&=\lambda^{[i]}\Ket{\psi^{[i]}} \label{GEVdefa} \\
\sum\limits_{k=0}^{n-1} \Braket{Q_j|I|Q_k} \alpha^{[i]}_k &=
  \lambda^{[i]} \sum\limits_{k=0}^{n-1} \Braket{ Q_j|Q_k} \alpha^{[i]}_k
  \label{GEVa} \\
 \psi^{[i]}(x)&=\sum\limits_{k=0}^{n-1} \alpha^{[i]}_k Q_k(x)
 \label{psiCa}
\end{align}
and solved.
Whereas the calculation of the moments
$\Braket{Q_m I}$, $\Braket{Q_m pI}$, $\Braket{Q_m \frac{dp}{dt}}$,
$\Braket{Q_m V\frac{dp}{dt}}$,
$\Braket{Q_m p}$ create no problem whatsoever,
an attempt to go beyond them
turned out to be problematic.
For example any second order derivative (e.g. $d^2p/dt^2$) cannot be obtained
directly from sample (\ref{dFSampleAll}) and,
in the same time, has singularities
when applying an integration by parts (\ref{momsDefinitionByParts}),
not to mention difficulties to formulate a boundary condition at $x=x_0$.

\begin{figure}[t]
  \includegraphics[width=8cm]{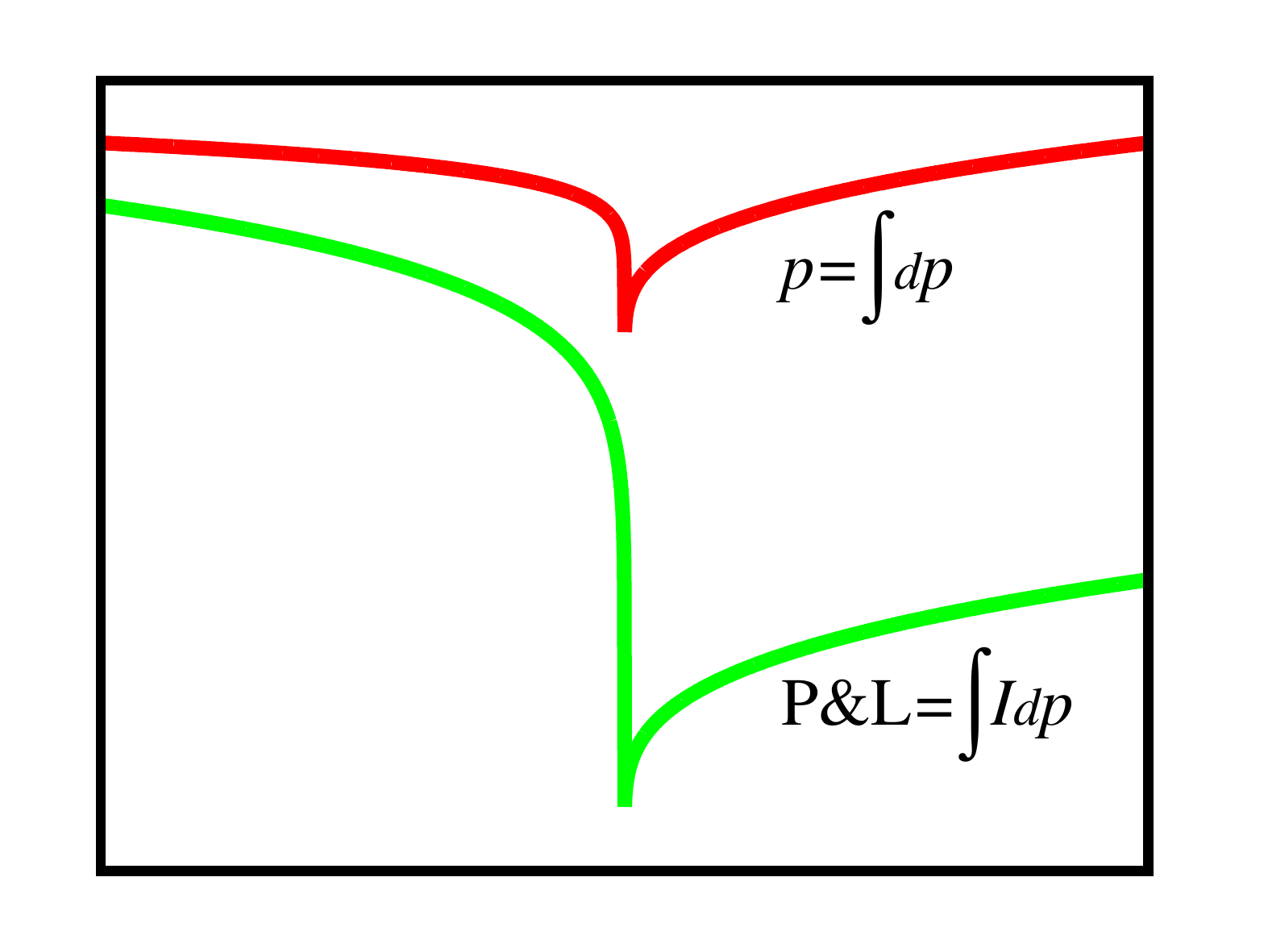}
  \caption{\label{PPnLopportunitiesDiff}
    A schematic example of price (the sum of price elementary changes $dp$) vs. possible P\&L
    (the sum of price changes multiplied by execution flow $Idp$).
    The second curve tells how much money one can be \textsl{potentially} made (or lost)
    on market movements taking into account traded volume capacity.
  }
\end{figure}

However, some of these characteristics are of great interest. The most important one
is $\Braket{Q_mI \frac{dp}{dt}}$.
The price $p=\int \frac{dp}{dt} dt=\int dp$ provides an information
of current market state, but little one about possible trading opportunities.
Assume at a given time interval $dt$ we have some specific constant
value of execution flow $I=dV/dt$ and some $dp/dt$.
During this $dt$ interval $dV=Idt$ shares were traded and the price change was $dp=\frac{dp}{dt}dt$.
How much money we potentially can make during this $dt$?
Buy at the beginning of $t$: $dV$ shares at $p$.
Sell at the end of $t+dt$: $dV$ shares at $p+dp$.
Total potential P\&L (assuming we can perfectly frontrun the market\footnote{
During every $dt$ we hold $dV$ shares, i.e. we always hold a position $S$ equals to execution flow $I=dV/dt$,
the $\mathrm{P\&L}=\int S(t)dp$, see \cite{2015arXiv151005510G}, Section ``P\&L operator and trading strategy''.})
is then $d\mathrm{P\&L}=Idp$.
The $\mathrm{P\&L}=\int I dp$
tells us how much money can be \textsl{potentially} made (or lost)
on market movements taking into account traded volume capacity.
This is the same sum of price changes as for regular price $p=\int dp$, but not all $dp$
are created equal.
If $dp$ occurred on a large execution flow -- it contributes more,
if on a small -- it contributes less\footnote{
The concept of $I\frac{dp}{dt}=\frac{dV}{dt}\frac{dp}{dt}$ is very different
from commonly studied
\href{https://en.wikipedia.org/wiki/Market_impact\#Market_impact_cost}{market impact}
concept
that is a price sensitivity to volume traded: $dp/dV$.
}
This creates a different way to study opportunities of market movements,
see Fig. \ref{PPnLopportunitiesDiff}.

The value of
$\Braket{Q_m I\frac{dp}{dt}}$ cannot be calculated directly
as well as it cannot be calculated using integration
by parts (\ref{momsDefinitionByParts})
in a general basis. However, if we change the basis and
vary basis functions
in the
basis of (\ref{GEVdefa}) eigenproblem
$\Ket{I\middle|\psi^{[i]}}=\lambda^{[i]}\Ket{\psi^{[i]}}$
an approximate solution can be obtained.
Consider the basis $\Ket{\psi^{[i]}}$, it is orthogonal
(\ref{normPsi}), (\ref{normPsiEV})
as:
\begin{subequations}
  \label{orthogonalPsiI}
  \begin{align}
    \delta_{jk}=\Braket{\psi^{[j]}|\psi^{[k]}} \\
    \lambda^{[j]}\delta_{jk}=\Braket{\psi^{[j]}|I|\psi^{[k]}}
  \end{align}
\end{subequations}
what gives $\Braket{\varphi|I|\psi^{[i]}}=\lambda^{[i]}\Braket{\varphi|\psi^{[i]}}$
for an arbitrary $\Ket{\varphi}$.
Then taking into account (\ref{momsDefinitionByParts}) and (\ref{ddtPsi2})
put $f=pI$ and $P^{last}=p(t_{now})$.  Obtain for matrix elements:
\begin{align}
  \Braket{\psi^{[j]}\left|\frac{dpI}{dt}\right|\psi^{[k]}}&=
  P^{last}I(t_{now})\omega(x_0)\psi^{[j]}(x_0)\psi^{[k]}(x_0) \nonumber\\
  &-\Braket{\mathrm{ED}(\psi^{[j]})|pI|\psi^{[k]}}
  -\Braket{\psi^{[j]}|pI|\mathrm{ED}(\psi^{[k]})}
  \label{momsDefinitionByPartsPsiIBasis}
\end{align}
These are  matrix elements of $\Braket{\psi^{[j]}\left|\frac{dpI}{dt}\right|\psi^{[k]}}$.
We are going to modify them to obtain
sought
matrix elements $\Braket{\psi^{[j]}\left|I\frac{dp}{dt}\right|\psi^{[k]}}$,
that should be zero when $p=const$. For this reason
the average in (\ref{momsDefinitionByPartsPsiIBasis})
should be reduced to an integral of an exact differential to be canceled
with out of integral term.
Taking into account (\ref{GEVdefa})
and (\ref{orthogonalPsiI})
one can see that
\begin{align}
  \Braket{\psi^{[j]}\left|I\frac{dp}{dt}\right|\psi^{[k]}}&=
  P^{last}\sqrt{\lambda^{[j]}\lambda^{[k]}}\omega(x_0)\psi^{[j]}(x_0)\psi^{[k]}(x_0) \nonumber\\
  &-\sqrt{\frac{\lambda^{[j]}}{\lambda^{[k]}}}\Braket{\mathrm{ED}(\psi^{[j]})|pI|\psi^{[k]}}
  -\sqrt{\frac{\lambda^{[k]}}{\lambda^{[j]}}}\Braket{\psi^{[j]}|pI|\mathrm{ED}(\psi^{[k]})}
  \label{momsDefinitionByPartsPsiIBasisModif}
\end{align}
satisfies exact differential condition. Then practically applicable expression is:
\begin{align}
  \Braket{\psi^{[j]}\left|I\frac{dp}{dt}\right|\psi^{[k]}}&=
  &\sqrt{\frac{\lambda^{[j]}}{\lambda^{[k]}}}\Braket{\mathrm{ED}(\psi^{[j]})|(P^{last}-p)I|\psi^{[k]}}
  +\sqrt{\frac{\lambda^{[k]}}{\lambda^{[j]}}}\Braket{\psi^{[j]}|(P^{last}-p)I|\mathrm{ED}(\psi^{[k]})}
  \label{momsDefinitionByPartsPsiIBasisModifPractical}
\end{align}
This is an expression of operator $\left\|I\frac{dp}{dt}\right\|$ in the basis
of (\ref{GEVdefa}).
The reason why we were able to obtain this explicit expression
is that we managed to combine differentiation of a product
(\ref{ddtPsi2}) with eigenvalues problem (\ref{GEVdefa})
 to write, when $p=const$, each matrix element as an
\href{https://en.wikipedia.org/wiki/Exact_differential}{exact differential}.
This approximation can also be viewed as operators multiplication factoring:
\begin{align}
  \Braket{\varphi|pI|\psi^{[i]}}&\approx
  \sum\limits_{k=0}^{n-1}
  \Braket{\varphi|p|\psi^{[k]}}\Braket{\psi^{[k]}|I|\psi^{[i]}}
  =\Braket{\varphi|p|\psi^{[i]}}\lambda^{[i]}
  \label{pIapprox}
\end{align}
what gives
\begin{align}
 &\mathrm{DP}_{jk}=\frac{1}{\lambda^{[k]}}\Braket{\mathrm{ED}(\psi^{[j]})|(P^{last}-p)I|\psi^{[k]}}
  +\frac{1}{\lambda^{[j]}}\Braket{\psi^{[j]}|(P^{last}-p)I|\mathrm{ED}(\psi^{[k]})}
  \label{DPjk}
\end{align}
and two approximations for $\|I\frac{dp}{dt}\|$:
\begin{align}
   \Braket{\psi^{[j]}\left|I\frac{dp}{dt}\right|\psi^{[k]}}&\approx
  \sqrt{\lambda^{[j]}\lambda^{[k]}}\mathrm{DP}_{jk}
  =\|I^{1/2}\|\cdot\|\mathrm{DP}\|\cdot\|I^{1/2}\|
  \label{DPwithITwoMultiplies} \\
  \Braket{\psi^{[j]}\left|I\frac{dp}{dt}\right|\psi^{[k]}}&\approx
  \|\mathrm{DP}\|\cdot\|I\|
  \label{DPwithI}
\end{align}
Using an approximation $\|I^{\beta}\frac{dp}{dt}\|\approx\|I^{\frac{\beta}{2}}|\frac{dp}{dt}|I^{\frac{\beta}{2}}\|$
the (\ref{DPwithITwoMultiplies}) result can be generalized to a $I$--power $\beta$,
however this result is not very accurate for $\beta$ other than $0$ or $1$,
since it is obtained from the $\Braket{Q_m pI}$ moments:
\begin{align}
   \Braket{\psi^{[j]}\left|I^{\beta}\frac{dp}{dt}\right|\psi^{[k]}}&\approx
  \left(\lambda^{[j]}\lambda^{[k]}\right)^{\frac{\beta}{2}}\mathrm{DP}_{jk}
  =\|I^{\beta/2}\|\cdot\|\mathrm{DP}\|\cdot\|I^{\beta/2}\|
  \label{momsDefinitionByPartsPsiIBasisModifPracticalAnyF}
\end{align}
This method of calculation is implemented in
\texttt{\seqsplit{com/polytechnik/freemoney/MatricesFromPI.java:getbQQIpowdpdtFromQQpi\_withLastP}};
for $\mathrm{DP}_{jk}$ see e.g.
\texttt{\seqsplit{com/polytechnik/freemoney/PFuture.java:bQQ\_DP}}.
A matrix obtained in $\Ket{\psi^{[i]}}$ basis can be converted to $Q_j$ basis
in a regular way:
\begin{align}
  \Braket{Q_j|f|Q_k} &= \sum\limits_{l,i,s,m=0}^{n-1} G_{jl} \alpha^{[i]}_l \Braket{\psi^{[i]}|f|\psi^{[s]}} \alpha^{[s]}_m G_{mk}
  \label{MatrFromBToQ} \\
  \Braket{\psi^{[j]}|f|\psi^{[k]}}
  &=\sum\limits_{s,m=0}^{n-1}\alpha^{[j]}_s \Braket{ Q_s|f|Q_m} \alpha^{[k]}_m
  \label{MatrFromQToB}
\end{align}
where Gram matrix $G_{jk}=\Braket{Q_j|Q_k}$, and $\alpha^{[i]}_k$ are
(\ref{GEVdefa}) eigenvectors in $Q_j$ basis, Eq. (\ref{psiCa}),
see \texttt{\seqsplit{com/polytechnik/utils/EVXData.java}}.
\footnote{
\scriptsize
A question arise whether
obtained matrix
$\Braket{Q_j\left|I^{\beta}\frac{dp}{dt}\right|Q_k}$, $j,k=0\dots n-1$
corresponds to a measure or not:
Whether it can be obtained
from some
$\Braket{Q_m I^{\beta}\frac{dp}{dt}}$ moments, $m=0\dots 2n-2$
by applying multiplication operator (\ref{multiplicationOperator})?
We have an algorithm to establish this fact,
see Theorem 3 of \cite{ArxivMalyshkinLebesgue}.
Numerical experiments show that this is almost the case.
A mismatch may be caused either by some numerical instability
or by some degeneracy in the problem.
A numerical instability is
very unlikely because (\ref{spurFromP})
holds exactly for both:
1). original $\Braket{Q_j\left|I^{\beta}\frac{dp}{dt}\right|Q_k}$ matrix
(\ref{DPwithITwoMultiplies})
and 2). the matrix obtained by applying multiplication operator
(\ref{multiplicationOperator})
to the moments $\Braket{Q_m I^{\beta}\frac{dp}{dt}}$
obtained from  Theorem 3 of  \cite{ArxivMalyshkinLebesgue}
applied to the original matrix
(\ref{DPwithITwoMultiplies}).
A mismatch between these two matrices
is observed starting with $n\ge 3$;
the difference is very small but clearly established numerically.
This degeneracy (if exists) does not create any problem
in calculation of the value of any observable
as the density matrix used has the form: a state since $\Ket{\psi^{[IH]}}$ spike.
}
The expression (\ref{DPwithITwoMultiplies})
is an approximation, similar (and slightly better) approximation
is a non--Hermitian matrix  (\ref{DPwithI})
as it has a single approximate product
compared with two in (\ref{DPwithITwoMultiplies}).

Different approximations can be obtained
using a general form of (\ref{pIapprox}):
\begin{align}
  \left\|fg\right\|&\approx
  \left\|f\right\|\cdot\left\|g\right\|
  \label{fgGeneral}
\end{align}
corresponding to an approximation
$\delta(x-y)\approx\sum_{j,k=0}^{n-1}Q_j(x)G^{-1}_{jk}Q_k(y)$.
We can obtain a result applying it to
other moments available directly from sample:
$\Braket{Q_m V \frac{dp}{dt}}$,
$\Braket{Q_m \frac{dp}{dt}}$
along with their derivatives
$\Braket{Q_m \frac{dI}{dt}}$,
$\Braket{Q_m \frac{dpI}{dt}}$,
$\Braket{Q_m \frac{d}{dt}V\frac{dp}{dt}}$,
and $\Braket{Q_m \frac{d^2p}{dt^2}}$
obtained from
integration by parts formula (\ref{momsDefinitionByParts})
with boundary conditions:
1). impact from the future  (\ref{iofuture})
and 2). zero $V$ at $t_{now}$;
traded volume $V$ is measured relatively $t_{now}$, it is negative
for past observations.
A few useful approximations of operator $\left\|I\frac{dp}{dt}\right\|$:
\begin{align}
    &\left\|\frac{dp}{dt}\right\| \cdot \left\|I\right\|  \label{withdPI}  \\
    &\left\|\frac{dpI}{dt}\right\| -
    \left\|p\right\| \cdot
    \left\|\frac{dI}{dt}\right\| \label{dPIwithP}\\
    &\left\|\frac{d}{dt}V \frac{dp}{dt}\right\| -  \left\|V\right\| \cdot \left\|\frac{d^2p}{dt^2}\right\|  \label{withPdI}    
\end{align}
Despite these operators being non-Hermitian
this creates no problem
as they are used only in
calculation of
$\mathrm{Spur}$  with Hermitian density matrix such as in (\ref{LAllTermsDelta}).
The approximation (\ref{withdPI})
uses directly sampled moments $\Braket{Q_m \frac{dp}{dt}}$,
it is a product
of two separately sampled operators with spikes,
nevertheless it gives very similar to  (\ref{DPwithI})
results, without  spurious artifacts;
sometimes, however, there is a difficulty to combine it
with $\|\frac{dpI}{dt}\|$ from (\ref{pIdiff}),
as in this case the moments from two different samplings are used together.
The approximation (\ref{dPIwithP}) uses
exact $\left\|\frac{dpI}{dt}\right\|$ and $\left\|\frac{dI}{dt}\right\|$
operators
matrix elements but the result is noisy.
The (\ref{withPdI}) also uses exact values of
$\left\|\frac{d}{dt}V \frac{dp}{dt}\right\|$
and
$\left\|\frac{d^2p}{dt^2}\right\|$ (obtained from (\ref{momsDefinitionByParts})
with zero boundary condition due to $V$ factor) operators,
but it is completely useless due to $d^2p/dt^2$ singularities;
however without the last term the  operator
$\left\|\frac{d}{dt}V \frac{dp}{dt}\right\|$
gives similar to 
$\left\|\frac{dpI}{dt}\right\|$
results. Two these operators should probably
be considered together as corresponding to $Vdp$ and $pdV$.
In sufficiently localizes states (e.g. $\Ket{\psi^{[IH]}}$) we typically have
$\Braket{\psi^{[IH]}|V \frac{dp}{dt}|\psi^{[IH]}}
\approx
\Braket{\psi^{[IH]}|V|\psi^{[IH]}}\Braket{\psi^{[IH]}|\frac{dp}{dt}|\psi^{[IH]}}$.

Another possible approach to obtain $\left\|I\frac{dp}{dt}\right\|$
that enters into (\ref{PEQIDelta}) together with $\left\|p\frac{dI}{dt}\right\|$
is to consider the operator $\left\|\frac{d}{dt}\frac{p}{I}\right\|$.
As above, let us take (\ref{momsDefinitionByPartsPsiIBasis})
and,
assuming that changes in $p$ are much smaller than changes in $I$,
modify it to obtain $\left\|\frac{d}{dt}\frac{p}{I}\right\|$
matrix elements:
\begin{align}
  \Braket{\psi^{[j]}\left|\frac{d}{dt}\frac{p}{I}\right|\psi^{[k]}}&\approx
  \frac{P^{last}}{I_0^F}\omega(x_0)\psi^{[j]}(x_0)\psi^{[k]}(x_0) \nonumber \\
    &-\frac{1}{{\lambda^{[k]}}^2}\Braket{\mathrm{ED}(\psi^{[j]})|pI|\psi^{[k]}}
    -\frac{1}{{\lambda^{[j]}}^2}\Braket{\psi^{[j]}|pI|\mathrm{ED}(\psi^{[k]})}
  \label{momsDefinitionByPartsPDivIBasisModifPractical}
\end{align}
This expression satisfies limit case conditions.
When $p=const$ the result
exactly equals to $\left\|\frac{d}{dt}\frac{1}{I}\right\|$,
when $I=const$ the result exactly equals to $\left\|\frac{dp}{dt}\right\|$,
and when $\|pI\|=\|I\|\cdot\|I\|$ the result is a differential of a constant.
With $I_0^F=\lambda^{[IH]}$ as per (\ref{iofuture}),
the (\ref{momsDefinitionByPartsPDivIBasisModifPractical})
also satisfies
$\mathrm{Spur}\left\|\frac{d}{dt}\frac{1}{I}\middle|\rho_{JIH}\right\|=0$,
the same as for 
$\mathrm{Spur}\left\|\frac{d}{dt}I\middle|\rho_{JIH}\right\|=0$.
See \texttt{\seqsplit{com/polytechnik/freemoney/MatricesFromPI.java:getbQQDtpDivIFromQQpi\_withLastP}}
for numerical implementation.
This approximation $\left\|\frac{d}{dt}\frac{p}{I}\right\|=
\left\|\frac{Idp/dt -pdI/dt}{I^2}\right\|$
can be tried as a proxy to $\left\|Idp/dt -pdI/dt\right\|$
in (\ref{PEQIDelta}), but the result is very poor.
The problem is that (\ref{momsDefinitionByPartsPDivIBasisModifPractical})
 has an extra common factor $1/I^2$.
 Similarly to
(\ref{DPjk})
we can modify it by $\lambda^{[j]}\lambda^{[k]}$ factor to remove $I^2$
from the denominator:
\begin{align}
  \left\|\psi^{[j]}\middle|I\frac{dp}{dt}-p\frac{dI}{dt}\middle|\psi^{[k]}\right\|&\approx
  \left\|I\right\| \cdot
  \left\|\frac{d}{dt}\frac{p}{I}\right\|
   \cdot\left\|I\right\| \label{IdpmPdiFromDPdivI} \\
  \Braket{\psi^{[j]}\left|I\frac{dp}{dt}-p\frac{dI}{dt}\right|\psi^{[k]}}&\approx
  \frac{P^{last}}{I_0^F}\omega(x_0)\lambda^{[j]}\lambda^{[k]}\psi^{[j]}(x_0)\psi^{[k]}(x_0) \nonumber \\
    &-\frac{\lambda^{[j]}}{\lambda^{[k]}}\Braket{\mathrm{ED}(\psi^{[j]})|pI|\psi^{[k]}}
  -\frac{\lambda^{[k]}}{\lambda^{[j]}}\Braket{\psi^{[j]}|pI|\mathrm{ED}(\psi^{[k]})}
  \label{momsDefinitionByPartsPDivIMultiplyI2BasisModifPractical}
\end{align}
but this creates a problem that the condition
$\mathrm{Spur}\left\|I\middle|\frac{d}{dt}\frac{1}{I}\middle|I\middle|\rho_{JIH}\right\|=0$
no longer holds, thus it cannot be applied in (\ref{PEQIDelta}).
One may try to adjust the value of $I_0^F$ in (\ref{momsDefinitionByPartsPDivIMultiplyI2BasisModifPractical})
to have this condition satisfied (put $p=const$ and take the Spur with $\|\rho_{JIH}\|$
of (\ref{momsDefinitionByPartsPDivIMultiplyI2BasisModifPractical}),
let it equals to zero; the values of
$I_0^F$ and ``adjusted'' $I_0^F$ are typically very close,
adjusted value is often slightly larger).
An important difference between 
$\left\|\frac{Idp/dt -pdI/dt}{I^2}\right\|$
and $\left\|Idp/dt -pdI/dt\right\|$
is that the first one is an exact differential (thus it's Spur with $\rho_{JIH}$
can be reduced to sub-differential expression in $\Ket{\psi^{[IH]}}$ state and the boundary term),
and the second one is not, thus it cannot be reduced to some observable
in $\Ket{\psi^{[IH]}}$ state.

The (\ref{DPwithI}),
(\ref{withdPI}),
and
(\ref{momsDefinitionByPartsPDivIMultiplyI2BasisModifPractical})
are approximations
that can be used for operator $\left\|I\frac{dp}{dt}\right\|$,
see \texttt{\seqsplit{com/polytechnik/freemoney/PFuture.java}}
for numerical implementation.

\section{\label{surrogateVolume}On Surrogate Volume}

\begin{figure}[t]

  \includegraphics[width=13cm]{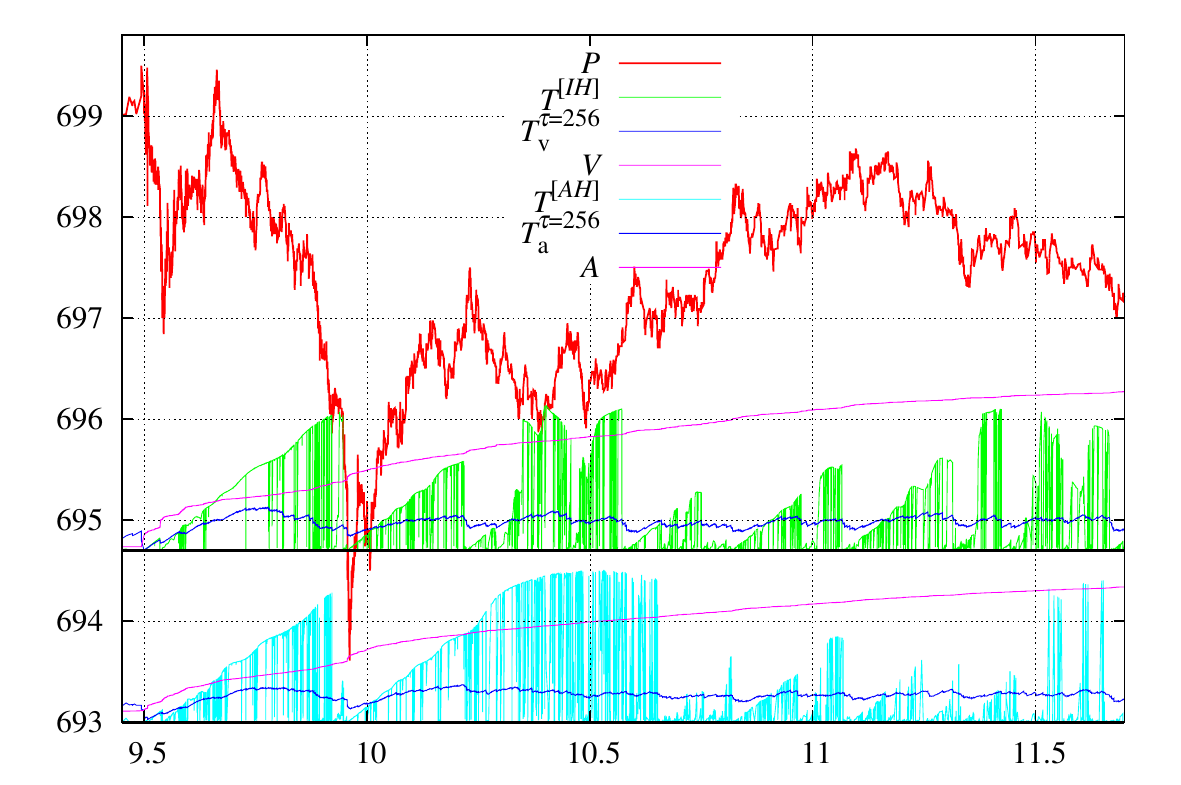}
  \caption{\label{TforVsurrogate}
    Averages: 
    $T^{[IH]}$ (\ref{TIHGEV})
    and regular moving average (\ref{TregularMovingAvergage})
    (dark blue) $T^{\tau}$ for $\tau=256$s;
    the values are multiplied by $10^{-3}$ and shifted up to fit the chart).
    The time-to now calculations with real volume (green) and surrogate volume (light blue).
    Real volume $V$ and surrogate volume $A$ are also presented (scaled by $10^{-6}$ and $5\cdot10^{-3}$ respectively) and shifted up to fit the chart).
 The AAPL stock on September, 20, 2012.
 The calculations in shifted Legendre basis with $n=12$.
  }
\end{figure}

Another question of interest is whether all the developed theory and software
of this work
can be used without trading volume available.
For a number of markets (such as: sovereign CDS, corporate fixed income, crypto exchanges, currency trading, etc.)
it is quite common to have
the price to be very accurate and available almost realtime, but the traded volume is either not
available at all or provided incorrectly (sometimes intentionally incorrectly).
In such a case there is
an option
to use the \textsl{absolute} value of price change as it were
the volume; the calculations are the same -- in (\ref{dFSampleAll})
instead of $df=V_l-V_{l-1}$ one can use
$da=|p_l-p_{l-1}|$. The only problem with this approach is that
market events without price change would not be taken into account
as for them $da=0$, hence the results will be less accurate.
However, our past experiments, see \cite{2015arXiv151005510G}, Fig. 6,
show that absolute value of price tick as a ``poor man volume''
often provides quite similar results.
A feed of ``all price ticks''
can be used as a surrogate volume.
For a liquid asset it typically takes a few minutes
for $\sum da$ to exceed $P^{last}-P^{0}=\sum dp$ in several orders of magnitude;
for the entire trading session in Fig. \ref{TforVsurrogate}
a typical maximal price change is about $5$, 
but the sum of all absolute price changes is about $500$;
total reported by NASDAQ ITCH \cite{itchfeed}
trading volume of this session is about $3\cdot 10^6$ shares.
As this $da$ sequence is all positive, we also tried it
with the market impact $dp/dV$ concept (that completely failed with actual price
change $dp$) in a hope that with this $da$ we may find
an identifiable limit for $dV/da$.
The result is also unsatisfactory: The $dV/da$ is very similar to $dV/dt$:
it fluctuates in orders of magnitude and clearly has no stable limit
at any time scale below 10 minutes (for US equity marker).
However the $da/dt$ is similar to $dV/dt$ and for liquid assets
can be used as a ``poor man volume'' $I=\frac{da}{dt}$
with the matrices $\Braket{Q_j|\frac{da}{dt}|Q_k}$ and
$\Braket{Q_j|Q_k}$ in (\ref{GEVa}).
See
\texttt{\seqsplit{com/polytechnik/freemoney/CommonlyUsedMoments.java:addObservationNoBasisShift}}
for calculation of surrogate volume moments.

\begin{figure}


  \includegraphics[width=14cm]{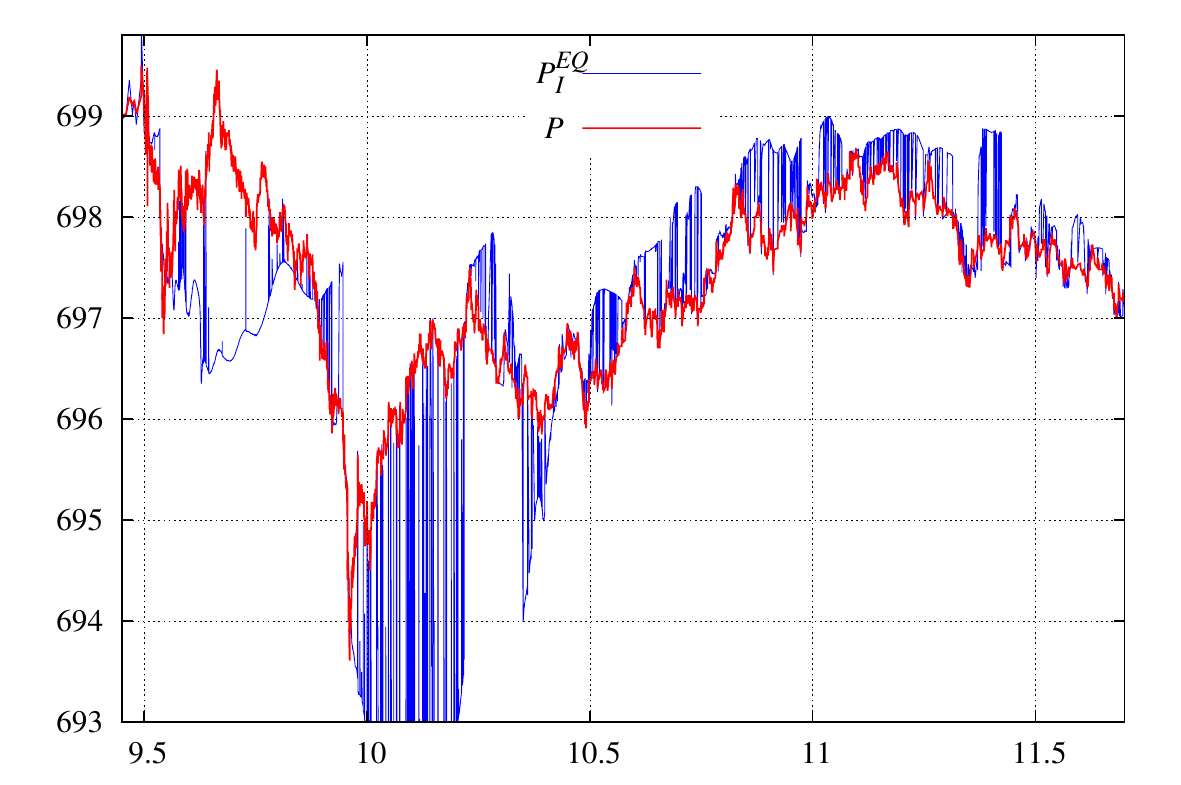}
  \includegraphics[width=14cm]{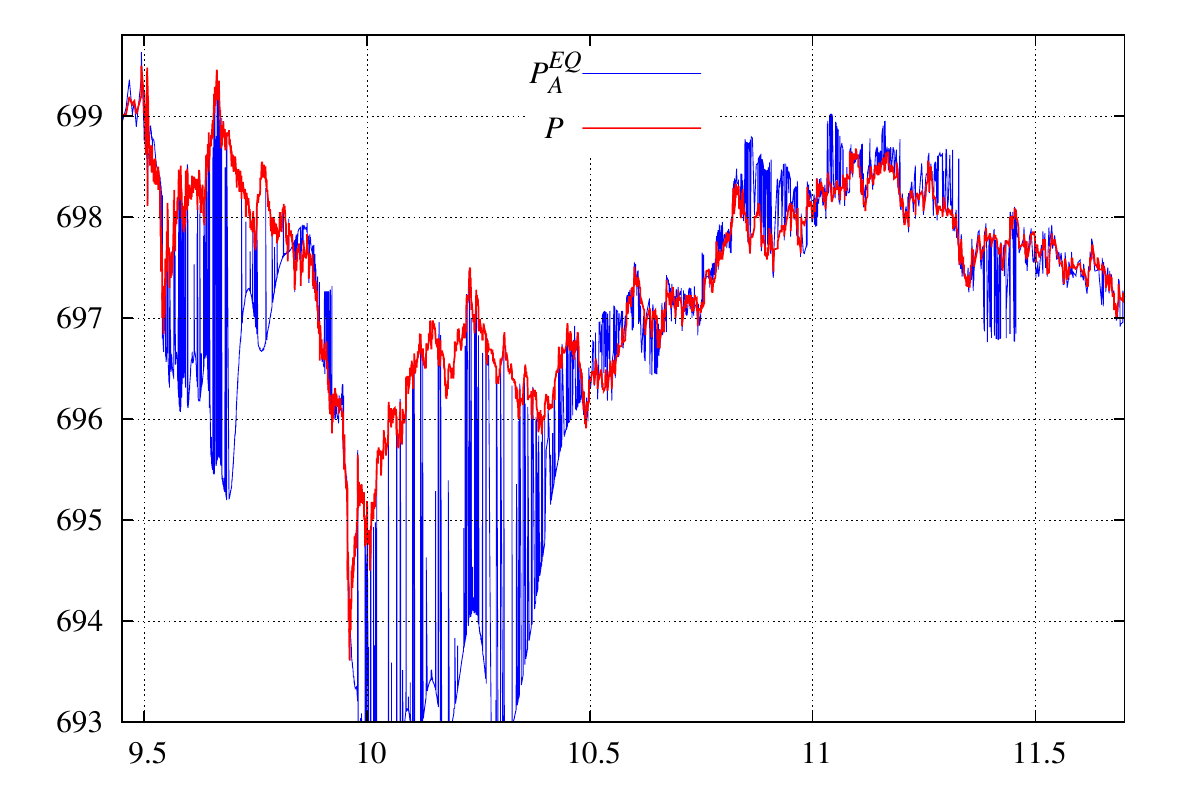}
  \caption{\label{PVforVsurrogate}
    The equilibrium price $P^{EQ}$ from (\ref{PEQV})
    for real volume $P^{EQ}_I$ and surrogate volume $P^{EQ}_A$.
  }  
\end{figure}

In Fig. \ref{TforVsurrogate} we present $T^{[IH]}$ (\ref{TIHGEV})
average (along with regular moving average $T^{\tau}$)
calculated for regular volume $dV$ and surrogate volume $da=|p_l-p_{l-1}|$.
One can see similar behavior of localization ``switches''.
However, surrogate volume states have some ``switches'' missed
and overall picture is less detailed.
In Fig. \ref{PVforVsurrogate} the price $P^{EQ}$ from (\ref{PEQV})
is presented. One can see similar, but less detailed picture.
This makes us to conclude that $da=|p_l-p_{l-1}|$
is a ``poor man volume''.

\section{\label{SoftwareDescription}Software Usage Description}
The software is written in java. As with
\cite{ArxivMalyshkinMuse} follow the steps:
\begin{itemize}
\item Install java 19 or later.
\item
  Download
  from \cite{polynomialcode}
NASDAQ ITCH data file
  \texttt{\seqsplit{S092012-v41.txt.gz}}, and
  the archive
\href{http://www.ioffe.ru/LNEPS/malyshkin/AMuseOfCashFlowAndLiquidityDeficit.zip}{\texttt{\seqsplit{AMuseOfCashFlowAndLiquidityDeficit.zip}}}
with the source code.
There is also an
\href{https://disk.yandex.ru/d/AtPJ4a8copmZJ?locale=en}{alternative location}
of these files.
\item Decompress and recompile the program:
\begin{verbatim}
unzip AMuseOfCashFlowAndLiquidityDeficit.zip
javac -g com/polytechnik/*/*java
\end{verbatim}

\item Run the command to test the program
\begin{verbatim}
java com/polytechnik/algorithms/TestCall_FreeMoneyForAll \
    --musein_file=dataexamples/aapl_old.csv.gz \
    --musein_cols=9:1:2:3 \
    --n=12 \
    --tau=256 \
    --measure=CommonlyUsedMomentsLegendreShifted \
    --museout_file=museout.dat
\end{verbatim}
Program parameters are:
\begin{itemize}
\item[] \verb+--musein_file=aapl.csv+ : Input tab--separated file
  with (time, execution price, shares traded) triples timeserie. The file
  is possibly \verb+gzip+-compressed.
\item[] \verb+--musein_cols=9:1:2:3+ : Out of total 9 columns of
  \verb+dataexamples/aapl_old.csv.gz+ file, take column \#1 as time (nanoseconds since midnight),
  \#2 (execution price), and \#3 (shares traded), column index is base 0.
\item[] \verb+--museout_file=museout.dat+ : Output file name is set to \texttt{\seqsplit{museout.dat}}.
\item[] \verb+--n=12+ : Basis dimension. Typical values are: 2 (for testing a concept),
or some value about $[4\dots 12]$ for more advance use.
\item[]  \verb+--tau=256+ : Exponent time (in seconds) for the measure used.
\item[] \verb+--measure=CommonlyUsedMomentsLegendreShifted+
  The measure. The values 
  \texttt{\seqsplit{CommonlyUsedMomentsLaguerre,CommonlyUsedMomentsMonomials}}
  correspond to Laguerre measure and \texttt{\seqsplit{CommonlyUsedMomentsLegendreShifted}}
  corresponds to shifted Legendre measures.
  The results of \texttt{\seqsplit{CommonlyUsedMomentsMonomials}} (uses $Q_k(x)=x^k$)
  should be identical to \texttt{\seqsplit{CommonlyUsedMomentsLaguerre}} (uses $Q_k(x)=L_k(-x)$),
  as the measure is the same and all the calculations are $Q_k(x)$--basis invariant (but numerical stability is worse for \texttt{\seqsplit{CommonlyUsedMomentsMonomials}}).
\end{itemize}
\item The results are saved in output file \verb+museout.dat+.
  Among the values of interest are the following:
  \begin{itemize}
  \item[] \verb+pFV.pv_average+ Regular moving average  $P^{\tau}$ from (\ref{pmovingaver}).
  \item[] \verb+pFV.Tv_average+ The moving average  $T^{\tau}$ from (\ref{TregularMovingAvergage}).
  \item[] \verb+pFV.totalVolume+ Total volume traded up to current tick.
  \item[] \verb+pFV.pv_M+ The value of $P^{[IH]}$ (\ref{PIHGEV}).
  \item[] \verb+pFV.Tv_M+ The value of  $T^{[IH]}$ (\ref{TIHGEV}), an indicator
    of localization of  $\Ket{\psi^{[IH]}}$.
  \item[] \verb+pFV.I.wH_squared+ The value of  $\Braket{\psi_0|\psi^{[IH]}}^2$ (\ref{SignalsToIgnore}),
    an indicator of applicability of any prediction.
  \item[] \verb+pFV.PEQV_from_M+ The value of predicted ``lagging price''
    $P^{EQ}$ from (\ref{PEQVFormula}) that includes
    both $P^{[IH]}$ and trending term, the best directional indicator we managed to obtain so far, see Fig. \ref{PVforVsurrogate}.
  \end{itemize}
\end{itemize}
The output includes two versions: calculated with ``actual'' volume
(prefixed with \verb+pFV.+) and calculated with ``surrogate volume''
of Appendix \ref{surrogateVolume} (prefixed with \verb+pFA.+).

\subsection{\label{SoftwareCodeStructure}Software Code Structure}
Provided software is located in several directories:
\begin{itemize}
\item \verb+com/polytechnik/utils/+ General basis utilities including my Radon-Nikodym approach \cite{malyshkin2019radonnikodym} to machine learning.
\item \verb+com/polytechnik/lapack/+ Ported to java
  \href{http://www.netlib.org/lapack/}{LAPACK} library.
\item \verb+com/polytechnik/lapack/+  NASDAQ ITCH \cite{itchfeed} parsing.
\item \verb+com/polytechnik/trading/+  Both: ``scaffolding'' for new ideas
  and a ``graveyard'' for old ones. Also contains unit tests.
  One can run all unit tests (takes >10 hours) as
\begin{verbatim}
java com/polytechnik/trading/QVM
\end{verbatim}
or three simple unit test of this paper algorithms:
{\scriptsize
\begin{verbatim}
java com/polytechnik/trading/PnLInPsiHstateLegendreShifted\$PnLInPsiHstateLegendreShiftedTest

java com/polytechnik/trading/PnLInPsiHstateLaguerre\$PnLInPsiHstateLaguerreTest

java com/polytechnik/trading/PnLInPsiHstateMonomials\$PnLInPsiHstateMonomialsTest
\end{verbatim}
}
\item \verb+com/polytechnik/freemoney/+ The code implementing the theory of this paper. Most noticeable are:
  \begin{itemize}
  \item[] \verb+com/polytechnik/freemoney/CommonlyUsedMoments.java+
    Calculate the moments from
    (Time, Execution Price, Shares Traded) sequence of transactions
    using direct sampling.
  \item[] \verb+com/polytechnik/freemoney/FreeMoneyForAll.java+
    A wrapper to calculate the matrices
    $\Braket{Q_j|f|Q_k}$ from sampled moments $\Braket{Q_mf}$.
    Secondary sampling of Appendix \ref{SecondarySampling}
    is also included.
  \item[] \verb+com/polytechnik/freemoney/PFuture.java+
    To calculate all the theory of this paper.
  \end{itemize}
\item \verb+com/polytechnik/algorithms/+ Drivers to call various algorithms.
\end{itemize}  

\section{\label{SecondarySampling}On Secondary Sampling}

When direct sampling (\ref{dFSampleAll}) of an observable
is not available
an advanced technique of ``secondary sampling''\cite{MalMuseScalp}
can be applied to calculate the moments of it.
An example. Assume on every tick
a moving average is calculated.
Then this \textsl{calculated} value is used as it were
a new observable, and the moving average of this new observable
is calculated.
For trivial cases
this gives nothing new: a moving average of a moving
average is a moving average with different weight
(for exponential moving average it is the moving
average with twice lower exponent time).
However, calculated quantity ``as it were an observable''
can be a characteristic that describes an immanent property
of the system. In \cite{MalMuseScalp} we applied
this technique to the maximal eigenvalue $\lambda^{[IH]}$ of
eigenproblem (\ref{GEVdef}); we
treated tick change in $\lambda^{[IH]}$ (calculated value, see Fig. \ref{ExampleIPsiH}) as it were
a change in
the observable $f_l=\lambda^{[IH]}\big|_{t_l}$
and calculated the moments with
$df_{l}=\lambda^{[IH]}\left|_{t_l}-\lambda^{[IH]}\right|_{t_{l-1}}$ as $\Braket{Q_mg\frac{df}{dt}}=\sum_{l=1}^{M}Q_m(x(t_l))g(t_l) \omega(t_l) df_{l}$.
The simplest application of this ``calculated observable''
is the sum of price changes
corresponding to positive $\lambda^{[IH]}$ changes
what gives the scalp price\cite{MalMuseScalp}:
\begin{align}
  {\cal P}&=\sum_{l=1}^{M}
  \begin{cases}p(t_l)-p(t_{l-1}) & \text{\enspace if \quad $\lambda^{[IH]}\left|_{t_l}-\lambda^{[IH]}\right|_{t_{l-1}} >0$}
    \\ 0 & \text{\enspace otherwise}
  \end{cases}
  \label{scalpPrice}
\end{align}
Normalization ${\cal P}(t_{now})=P^{last}$ is typically used for scalp price.
Regular price corresponds to no condition on $\lambda^{[IH]}$.
Whereas actual computation is performed
in a single pass by highly optimized code
using recurrent relation for moments
and in-place calculation of the value to be used as a ``new observable'',
for understanding the concept
one may think about secondary sampling as having two-passes for
input timeserie: first --- scan all timeserie observations
and build a ``new observable'' for every timeserie point read; second ---
 scan this timeserie once again treating the value calculated on the
first pass as it were a regular observable.
This ``secondary sampling'' approach
greatly extends the types of observable that can be studied.
However, while being very powerful
in calculation of moments that otherwise
are not approachable at all,
it has difficulties in interpretation of the results.

The implementation of this technique is available in
\texttt{\seqsplit{com/polytechnik/freemoney/CommonlyUsedMoments.java}}.
The methods \texttt{\seqsplit{updateWithSingleObservation}}
recurrently adjusts the basis and adds calculated contributions
corresponding to regular measures $dP$, $dV$, $dA$ as in (\ref{dFSampleAll}).
After this call all regular moments become available,
and we have an option
to calculate the value of some ``secondary'' observable from them, such as $\lambda^{[IH]}$.
When the calculation is completed --- the method
\texttt{\seqsplit{addIHObservationSecondarySampling(double\, IH)}}
can be called. It, in addition to the moments already available from
\texttt{\seqsplit{updateWithSingleObservation}},
calculates, as in (\ref{dFSampleAll}), three other moments corresponding to the measure
$df_l=\texttt{IH}_l-\texttt{IH}_{l-1}$, specifically:
$\Braket{Q_m \frac{df}{dt}}$,
$\Braket{Q_m p\frac{df}{dt}}$,
and
$\Braket{Q_m \frac{dpf}{dt}}$.
The \texttt{IH} can be a calculated characteristic of various meaning,
and
the moments of this characteristic are now obtained as it were
a regular observable. This technique is also very convenient
for unit tests of moments calculation
by using regular observable as \texttt{IH}.

\section{\label{dIprojections}On Separation of States Based On $dI/dt$ Sign}
The value of future execution flow $I_0^{F}$ (\ref{iofuture})
allows us to obtain $\|dI/dt\|$ operator's matrix elements.
Using
integration by parts (put $p=const$ in (\ref{momsDefinitionByPartsPsiIBasis}))
obtain:
\begin{align}
  \Braket{\psi^{[j]}\left|\frac{dI}{dt}\right|\psi^{[k]}}&=
  I_0^{F}\omega(x_0)\psi^{[j]}(x_0)\psi^{[k]}(x_0)
  -\Braket{\mathrm{ED}(\psi^{[j]})|I|\psi^{[k]}}
  -\Braket{\psi^{[j]}|I|\mathrm{ED}(\psi^{[k]})}
  \label{momsByPartsPsiIBasisdIdt}
\end{align}
This operator's matrix elements cannot be obtained directly from sample (\ref{dFSampleAll}),
however the knowledge of the impact from the future
allows us to apply an integration by parts.
Actually this is
the only operator for which integration by parts gives exact answer;
for other operators (e.g $\|d^2p/dt^2\|$)
the boundary value at $x_0$ is not known and
matrix elements are typically obtained ``within a boundary term''.
Only having determined
the exact value of $I_0^{F}$ (that includes
both:
an ``impact from the past''
and
an ``impact from the future'')
it is possible to have
the $\|dI/dt\|$ matrix elements that are accurate enough
to consider an eigenproblem:
\begin{align}
  \Ket{\frac{dI}{dt}\middle|\psi_{dI}^{[i]}}
  &=\lambda_{dI}^{[i]}\Ket{\psi_{dI}^{[i]}}
  \label{dIdtEigenproblem}
\end{align}
Other than $I_0^{F}=\lambda^{[IH]}$ (\ref{iofuture})  values
can be used in the boundary term of (\ref{momsByPartsPsiIBasisdIdt}),
see \cite{MalMuseScalp} ``Appendix E: \textsl{On calculation of $dI/dt$ operator matrix elements from operator $I$}'' for a list of reasonable options for $I_0^{F}$.
The concept introduced in \cite{MalMuseScalp} is to treat
\textsl{low $I$ $\to$ high $I$} and \textsl{high $I$ $\to$ low $I$}
transitions \textbf{separately},
as they lead to a very different price behavior.
These $I$-transitions correspond to $dI/dt$ derivative of different signs;
corresponding operator $\|dI/dt\|$ always has eigenvalues $\lambda_{dI}^{[i]}$
of different signs: $\Braket{\psi^{[IH]}|\frac{dI}{dt}|\psi^{[IH]}}=0$.
Let us split the entire $\Ket{\psi}$ space into direct
sum of two subspaces\footnote{
Here we use $\|dI/dt\|$ matrix elements (\ref{momsByPartsPsiIBasisdIdt}) that are calculated
from directly sampled $\|I\|$ moments. An alternative
is to split the states according to $dI/dt$ or $d\lambda^{[IH]}/dt$ sign
using the ``secondary sampling'' of \cite{MalMuseScalp}.
The simplest example of it's application is
the scalp price ${\cal P}$ (\ref{scalpPrice})
that takes into account only ``important'' price changes
(regular price is the sum of all price changes).
}.
Construct two projection operators:
\begin{align}
  \|\Pi_{dI+}\|&=\sum\limits_{i:\,\lambda_{dI}^{[i]}>0} \Ket{\psi_{dI}^{[i]}}\Bra{\psi_{dI}^{[i]}}
  \label{dI+} \\
  \|\Pi_{dI-}\|&=\sum\limits_{i:\,\lambda_{dI}^{[i]}\le 0} \Ket{\psi_{dI}^{[i]}}\Bra{\psi_{dI}^{[i]}}
  \label{dI-} \\
  \|\mathds{1}\|&=\|\Pi_{dI+}\|+\|\Pi_{dI-}\| \label{sumsubspaces}
\end{align}
This transform can be considered
as eigenvalues adjustment technique\cite{gsmalyshkin2017comparative}
where the eigenvalues (not the eigenvectors!)
are adjusted for an effective identification of weak hydroacoustic signals.
The $\|\Pi_{dI+}\|$ can be viewed as $\|dI/dt\|$ operator with all negative eigenvalues set to
$0$ and all positive eigenvalues set to $1$; the same with $\|\Pi_{dI-}\|$ for opposite sign.
This technique is most easy to implement in (\ref{dIdtEigenproblem}) basis (where $\|dI/dt\|$ is diagonal),
then to convert obtained projection operators
back to the basis used applying (\ref{MatrFromBToQ}).
Alternatively one can convert all the matrices $\|I\|$, $\|pI\|$, $\|V\frac{dp}{dt}\|$, $\|\frac{d pI}{dt}\|$, $\|\frac{d}{dt}V\frac{dp}{dt}\|$,
$\left\|\rho\right\|$ to the basis of (\ref{dIdtEigenproblem}) eigenproblem applying (\ref{MatrFromQToB}).
All the results will be identical as the theory
is gauge invariant\cite{malyshkin2019radonnikodym}.
With projection operators (\ref{dI+}) and (\ref{dI-})
any density matrix average can be written in the form:
\begin{align}
  \mathrm{Spur}\left\|f\middle|\rho\right\|&=
  \mathrm{Spur}\left\|f\middle|\Pi_{dI+}\middle|\rho\right\|
  +  \mathrm{Spur}\left\|f\middle|\Pi_{dI-}\middle|\rho\right\|
  \label{splitProj}
\end{align}
This split allows us to separate an average of $f$ in density matrix $\|\rho\|$
state to the ones corresponding to positive and negative $dI/dt$.

First candidates on application of this technique are
the terms from Total Lagrangian action (\ref{STotal})
where the operators are calculated in the state of
$\left\|\rho_{JIH}\right\|$ density matrix.
Every $\mathrm{Spur}$ that enter into the $\Delta$
expression (\ref{PEQTotalDelta})
can be split into $dI/dt$ contributions of different signs.
There are several implementation of this technique, e.g.
\texttt{\seqsplit{com/polytechnik/freemoney/SplitdIdt.java}}
and several others. The result, however, is not that great and
this projection operators approach requires more research to be performed.

The property that requires special attention is that
while the density matrix $\|\rho_{JIH}\|$ is obtained
from the polynomial ${\psi^{[IH]}}^2(x)$ with $J\left({\psi^{[IH]}}^2\right)$
transform (\ref{wfPartsdef}),
and all the average relations hold exactly,
the density matrix
itself may not have all the eigenvalues positive.
This creates no problem with the total $\mathrm{Spur}$
but sometimes lead to spurious artifacts when combined with projection operators;
the effect, however, is small.
These small but negative eigenvalues of the density matrix,
``\href{https://en.wikipedia.org/wiki/Hermann_Minkowski}{Hermann Minkowski}-style space'', also require additional research.

\subsection{\label{IoperarorProjections}Execution Flow Based Eigenvalues Adjustment Example}
In the Appendix above we considered projection operators (\ref{sumsubspaces})
to ``split'' $\|I\|$ or $\|pI\|$
 based on some other operator spectrum, e.g. $\left\|\frac{dI}{dt}\right\|$.
To demonstrate a simplified example of this 
 eigenvalues adjustment technique let us apply it to the operator $\|I\|$.
Consider the state ``since $\Ket{\psi^{[IH]}}$ till now''
$\|\rho_{JIH}\|$ and a trading strategy: buy at execution flow below
aggregated execution flow $\frac{V_{IH}}{T_{IH}}$
with (\ref{PlastdVdT}) and (\ref{PlastConst}), 
and sell above it.
The P\&L position changes $dS$, see \cite{2015arXiv151005510G} Section ``P\&L operator and trading strategy'', is:
\begin{align}
  dS&=\left(I-\frac{V_{IH}}{T_{IH}}\right)dt \label{dSIpos}
\end{align}
Then the constraint $0=\int dS$ is satisfied:
\begin{align}
  0&=\mathrm{Spur}\left\|\frac{dS}{dt}\middle|\rho_{JIH}\right\|
  \label{dSconstraint}
\end{align}
and, for this $dS$, the  $\mathrm{P\&L}=-\int pdS$ can be calculated:
\begin{align}
  \mathrm{P\&L}&=-\mathrm{Spur}\left\|\frac{pdS}{dt}\middle|\rho_{JIH}\right\|
  \label{PdSPnL}
\end{align}
If we want to consider (\ref{PdSPnL}) as a superposition of
$\Ket{\psi^{[i]}}$ states, introduce an operator:
\begin{align}
  \Pi&=\sum\limits_{i=0}^{n-1}\Ket{\psi^{[i]}}\left(1-\frac{V_{IH}}{T_{IH}}\frac{1}{\lambda^{[i]}}\right)\Bra{\psi^{[i]}}
  \label{projPiIvalue}\\
  \Pi&=\Pi_{+}+\Pi_{-} \label{projPiIvalueSum}\\
  \Pi_{+}&=\sum\limits_{i:\,\lambda^{[i]}>\frac{V_{IH}}{T_{IH}}}^{n-1}\Ket{\psi^{[i]}}\left(1-\frac{V_{IH}}{T_{IH}}\frac{1}{\lambda^{[i]}}\right)\Bra{\psi^{[i]}} \label{projPiIvalue+}\\
  \Pi_{-}&=\sum\limits_{i:\,\lambda^{[i]}\le\frac{V_{IH}}{T_{IH}}}^{n-1}\Ket{\psi^{[i]}}\left(1-\frac{V_{IH}}{T_{IH}}\frac{1}{\lambda^{[i]}}\right)\Bra{\psi^{[i]}} \label{projPiIvalue-} 
\end{align}
The operator $\|\Pi\|$ is actually the operator $\|I\|$
but with the eigenvalues $1-\frac{V_{IH}}{T_{IH}}\frac{1}{\lambda^{[i]}}$
instead of the $\lambda^{[i]}$\footnote{
The operator $\|\Pi\|$ may not correspond to a measure,
i.e. it does not necessary correspond to $m=0\dots 2n-2$
moments $\Braket{Q_m\Pi}$ from which to obtain $\Braket{Q_j|\Pi|Q_k}$
using multiplication operator (\ref{multiplicationOperator}).
}.
Then  (\ref{dSconstraint}) and (\ref{PdSPnL}) become (\ref{dSconstraintPi}) and (\ref{PdSPnLPi}) respectively:
\begin{align}
  0&= \mathrm{Spur}\left\|I\middle|\Pi\middle|\rho_{JIH}\right\|
  =\mathrm{Spur}\left\|I\middle|\Pi_{+}+\Pi_{-}\middle|\rho_{JIH}\right\|
  \label{dSconstraintPi} \\
  \mathrm{P\&L}&=-\mathrm{Spur}\left\|pI\middle|\Pi\middle|\rho_{JIH}\right\|
  =-\mathrm{Spur}\left\|pI\middle|\Pi_{+}+\Pi_{-}\middle|\rho_{JIH}\right\|
  \label{PdSPnLPi}
\end{align}
From these operators $\|\Pi_{+}\|$ and $\|\Pi_{-}\|$ one can obtain
``equilibrium prices''
$P^{\pm}=\frac{\mathrm{Spur}\left\|pI\middle|\Pi_{\pm}\middle|\rho_{JIH}\right\|}
{\mathrm{Spur}\left\|I\middle|\Pi_{\pm}\middle|\rho_{JIH}\right\|}$ and etc.
The result is similar to the technique of ``extra volume'' $\widetilde{V}$
and $P^*$ of the Appendix \ref{doubleIntegration} below;
no advancing information we managed to obtain from (\ref{projPiIvalue}).

\section{\label{doubleIntegration}On The States Of Double Integration}
The density matrix state $\rho_{JIH}$ (\ref{rhoJIH})
was obtained from the pure state of maximal execution flow
$\Ket{\psi^{[IH]}}$ by applying $J(\cdot)$ transform (\ref{wfPartsdef})
to the polynomial ${\psi^{[IH]}}^2(x)$,
a variant of integration by parts:
\begin{align}
  &\mathrm{Spur}\left\|\frac{df}{dt}\middle|\rho_{JIH}\right\|=
  f\Big|_{x0}\rho_{JIH}(x_0)\omega(x_0)
  -\Braket{\psi^{[IH]}\left|f\right|\psi^{[IH]}}=
  f\Big|_{x0}-\Braket{\psi^{[IH]}\left|f\right|\psi^{[IH]}}
  \label{proJIHparts} \\
  &\sum\limits_{j,k,l,m=0}^{n-1}Q_j(x_0)G_{jk}^{-1}{\rho_{JIH}}_{kl}G_{lm}^{-1}Q_m(x_0)=1
  \label{proJHnorm}
\end{align}
with (\ref{proJHnorm}) due to $1=\Braket{\psi^{[IH]}|\psi^{[IH]}}$
normalizing and $1=\omega(x_0)$ due to basis choice obtain familiar
``integration by parts'' relation (\ref{JD1}).
The $\left\|\rho_{JIH}\right\|$ density matrix
allows us to calculate ``execution-flow''-related values from
$\left\|\frac{dI}{dt}\right\|$ operator.
We already used this relation (a special case of (\ref{spurFromP}))
to calculate e.g.
\begin{align}
  \mathrm{Spur}\left\|\frac{dp}{dt}\middle|\rho_{JIH}\right\|  &=
  P^{last}-\Braket{\psi|p|\psi} \label{PlastP}\\
  \mathrm{Spur}\left\|\frac{dpI}{dt}\middle|\rho_{JIH}\right\| &=
  P^{last}I_0^F-\Braket{\psi^{[IH]}\left|pI\right|\psi^{[IH]}}
  \label{PlastdPI} \\
  \mathrm{Spur}\left\|\frac{dI}{dt}\middle|\rho_{JIH}\right\| &=
  I_0^F-\Braket{\psi^{[IH]}\left|I\right|\psi^{[IH]}}
  \label{PlastdI} \\
  \mathrm{Spur}\left\|I\middle|\rho_{JIH}\right\| &= V_{IH}=
  V^{last}-\Braket{\psi^{[IH]}\left|V\right|\psi^{[IH]}}
  \label{PlastdVdT} \\
  \mathrm{Spur}\left\|\rho_{JIH}\right\| &= T_{IH}=
  T^{last}-\Braket{\psi^{[IH]}\left|T\right|\psi^{[IH]}}
   \label{PlastConst}
\end{align}
The (\ref{PlastdPI}) corresponds to (\ref{momsDefinitionByPartsPsiIBasis}),
(\ref{PlastdI}) with boundary condition (\ref{iofuture})
gives $\mathrm{Spur}\left\|\frac{dI}{dt}\middle|\rho_{JIH}\right\|=0$
we used in (\ref{SI}),
(\ref{PlastdVdT}) and (\ref{PlastConst})
are traded volume and time since $\Ket{\psi^{[IH]}}$ spike till ``now'';
typically we use normalizing $V^{last}=0$ and $T^{last}=0$.

\begin{figure}
  \includegraphics[width=16cm]{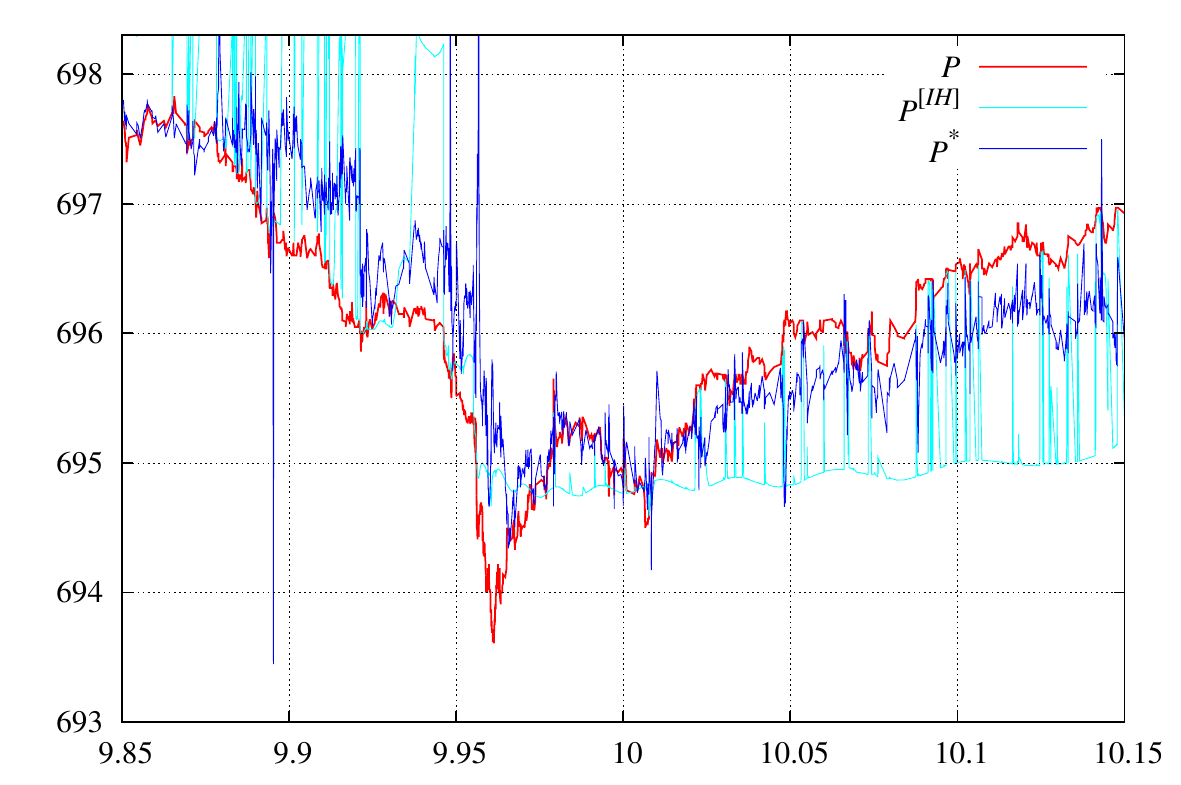}
  \caption{\label{PEQIVJJ}
    Price $P$, price $P^{[IH]}$ (\ref{PIHGEV}),
    and $P^*$ (\ref{PEQIDeltaJJ})
    for
 AAPL stock on September, 20, 2012.
 The calculations in shifted Legendre basis with $n=12$ and $\tau$=256sec.
  }
\end{figure}

Now consider a density matrix $\rho_{JJIH}$ (\ref{rhoJJIH})
obtained from the pure state of maximal execution flow
$\Ket{\psi^{[IH]}}$ by applying $J(\cdot)$ transform (\ref{wfPartsdef})
to the polynomial ${\psi^{[IH]}}^2(x)$ \textsl{twice}.
This density matrix corresponds
to integration by parts performed twice.
Obtain from (\ref{JD2}):
\begin{align}
  \mathrm{Spur}\left\|\frac{dI}{dt}\middle|\rho_{JJIH}\right\| &=
  I_0^F\mathrm{Spur}\left\|\rho_{JIH}\right\|
  -
  \mathrm{Spur}\left\|I\middle|\rho_{JIH}\right\|
  =I_0^F T_{IH}-V_{IH}
  \label{Itwice} \\
   \mathrm{Spur}\left\|\frac{dpI}{dt}\middle|\rho_{JJIH}\right\| &=
  I_0^FP^{last}\mathrm{Spur}\left\|\rho_{JIH}\right\|
  -
  \mathrm{Spur}\left\|pI\middle|\rho_{JIH}\right\|
  \label{PlastdItwice}
\end{align}
The $\left\|\rho_{JJIH}\right\|$ density matrix
allows us to calculate ``volume''-related values from $\left\|\frac{dI}{dt}\right\|$ operator.
One of the major results of this paper is
established in 
Section \ref{ExecutionFlowBasedDynamics}
fact that in $\left\|\rho_{JIH}\right\|$ state the values
of operators
$\left\|p\frac{dI}{dt}\right\|$
and
$\left\|I\frac{dp}{dt}\right\|$
are very close and their difference (if exists)
gives future price (\ref{PEQI}).
Let us consider these operators
not in $\left\|\rho_{JIH}\right\|$ state, but instead
in the state $\left\|\rho_{JJIH}\right\|$.
An important difference from  $\left\|\rho_{JIH}\right\|$ state
is that the condition
of $\left\|\frac{dI}{dt}\right\|$
being zero in $\left\|\rho_{JJIH}\right\|$ state no longer holds,
as execution flow $I=dV/dt$ and aggregated execution flow $V_{IH}/T_{IH}$
are different in $\Ket{\psi^{[IH]}}$ state. The (\ref{Itwice})
requires an ``extra volume'' $\widetilde{V}$
\begin{align}
  \widetilde{V}&=\left(I_0^F-\frac{V_{IH}}{T_{IH}}\right)T_{IH}
  \label{extraVol}
\end{align}
to obtain proper value of operator $\left\|p\frac{dI}{dt}\right\|$;
an alternative is to use eigenvalues adjustment technique
of Appendix \ref{IoperarorProjections} above.
Using (\ref{PEQIDelta}) obtain
\begin{align}
  \Delta&=\mathrm{Spur}\left\|I\frac{dp}{dt}\middle|\rho_{JJIH}\right\|-\mathrm{Spur}\left\|p\frac{dI}{dt}\middle|\rho_{JJIH}\right\| +\widetilde{V}P^*\nonumber \\
&=2\mathrm{Spur}\left\|I\frac{dp}{dt}\middle|\rho_{JJIH}\right\|
  -\mathrm{Spur}\left\|\frac{dpI}{dt}\middle|\rho_{JJIH}\right\|
  +\widetilde{V}P^*
    \label{PEQIDeltaJJ}
\end{align}
The difference from (\ref{PEQIDelta}) is that there is
an extra term $\widetilde{V}P^*$ caused by the difference
between $I_0^F$ and $\frac{V_{IH}}{T_{IH}}$.
We can consider $\Ket{V|\psi}=\lambda\Ket{T|\psi}$
providing equal execution flow and aggregated execution flow,
see \cite{MalMuseScalp}: ``Appendix C: The state of maximal aggregated execution flow $V/T$'', but these states gives little improvement.
Consider instead a simplistic approach: select the value of $P^*$ that makes
$\Delta$ equals to zero:
\begin{align}
  P^{*}&=-\,\frac{1}{\widetilde{V}}\left\lgroup
  2\mathrm{Spur}\left\|I\frac{dp}{dt}\middle|\rho_{JJIH}\right\|
  -\mathrm{Spur}\left\|\frac{dpI}{dt}\middle|\rho_{JJIH}\right\|
\right\rgroup
\end{align}
In Fig. \ref{PEQIVJJ} the $P^*$ is presented.
We see no ``advancing'' property as it is for (\ref{PEQI}),
this is an indicator of ``lagging'' type.
This makes us to conclude that the state
$\rho_{JJIH}$
while it has a number of interesting properties to research,
does not immediately provide
and ``advancing'' indicator. The $\rho_{JIH}$ is probably the
only state in which 
$\left\|p\frac{dI}{dt}\right\|$
and
$\left\|I\frac{dp}{dt}\right\|$
operators 
are very close and their difference (if exists)
gives future price (\ref{PEQI}).

\bibliography{LD}

\end{document}